\documentclass[11pt,english,onecolumn]{IEEEtran}
\usepackage[T1]{fontenc}
\usepackage[latin9]{inputenc}
\usepackage{color}
\usepackage{units}
\usepackage{amsthm}
\usepackage{amsmath}
\usepackage{amssymb}
\usepackage{mathdots}
\usepackage{setspace}
\PassOptionsToPackage{version=3}{mhchem}
\usepackage{mhchem}
\usepackage{esint}
\makeatletter

\providecommand{\tabularnewline}{\\}

\theoremstyle{plain}
\newtheorem{thm}{\protect\theoremname}
\theoremstyle{plain}
\newtheorem{conjecture}[thm]{\protect\conjecturename}
\theoremstyle{remark}
\newtheorem{rem}[thm]{\protect\remarkname}
\theoremstyle{definition}
\newtheorem{defn}[thm]{\protect\definitionname}
\theoremstyle{plain}
\newtheorem{lem}[thm]{\protect\lemmaname}
\theoremstyle{plain}
\newtheorem{cor}[thm]{\protect\corollaryname}

\usepackage{hyperref}
\usepackage{breqn}
  
 \DeclareMathOperator*{\argmin}{arg\,min}

\allowdisplaybreaks

\global\long\def\s[#1]{\mathrm{\scriptsize #1}}
\global\long\def\st[#1]{\mathrm{\scriptsize #1}}

\global\long\def\P{\mathbb{P}}
\global\long\def\E{\mathbb{E}}

\global\long\def\m{\mathsf{M}}
\global\long\def\l{\mathsf{L}}
\global\long\def\maj{\mathsf{Maj}}
\global\long\def\dic{\mathsf{Dict}}
\global\long\def\b{\mathsf{b}}

\global\long\def\binent{\mathsf{h_b}}
\global\long\def\bindiv{\mathsf{d_b}}
\global\long\def\Hamw{\mathsf{W_H}}

\global\long\def\dfn{:=}

\global\long\def\trre[#1,#2]{\overset{{\scriptstyle (#2)}}{#1}} % transition explained with reason

\author{Nir Weinberger, \IEEEmembership{Student Member, IEEE,} and Ofer Shayevitz, \IEEEmembership{Senior Member, IEEE}}

\makeatother

\usepackage{babel}
\providecommand{\conjecturename}{Conjecture}
\providecommand{\corollaryname}{Corollary}
\providecommand{\definitionname}{Definition}
\providecommand{\lemmaname}{Lemma}
\providecommand{\remarkname}{Remark}
\providecommand{\theoremname}{Theorem}

\begin{document}

\title{On the Optimal Boolean Function for Prediction under Quadratic Loss%
\thanks{The work of the first author was supported by the Gutwirth scholarship
for Ph.D. students of the Technion, Israel Institute of Technology.
The work of the second author was supported by an ERC grant no. 639573,
and an ISF grant no. 1367/14. The material in this paper was presented
in part at the IEEE International Symposium on Information Theory
(ISIT), Barcelona, Spain, July 2016. %
}}

\maketitle
\renewcommand\[{\begin{equation}}
\renewcommand\]{\end{equation}}
\begin{abstract}
Suppose $Y^{n}$ is obtained by observing a uniform Bernoulli random
vector $X^{n}$ through a binary symmetric channel. Courtade and Kumar
asked how large the mutual information between $Y^{n}$ and a Boolean
function $\b(X^{n})$ could be, and conjectured that the maximum is
attained by a dictator function. An equivalent formulation of this
conjecture is that dictator minimizes the prediction cost in a sequential
prediction of $Y^{n}$ under \emph{logarithmic loss}, given $\b(X^{n})$.
In this paper, we study the question of minimizing the sequential
prediction cost under a different (proper) loss function -- the \emph{quadratic
loss}. In the noiseless case, we show that majority asymptotically
minimizes this prediction cost among all Boolean functions. We further
show that for weak noise, majority is better than dictator, and that
for strong noise dictator outperforms majority. We conjecture that
for quadratic loss, there is no single sequence of Boolean functions
that is simultaneously (asymptotically) optimal at all noise levels.\end{abstract}

\section{Introduction and Problem Statement}

Let $X^{n}\in\{0,1\}^{n}$ be a uniform Bernoulli random vector,%
\footnote{As customary, upper case letters will denote random variables/vectors,
and their lower case counterparts will denote specific values that
they take.%
} and let $Y^{n}$ be the result of passing $X^{n}$ through a memoryless
binary symmetric channel (BSC) with crossover probability $\alpha\in[0,\frac{1}{2}]$.
Recently, Courtade and Kumar conjectured the following: 
\begin{conjecture}[\cite{Boolean_conjecture}]
\label{conj: Boolean function conjecture} For any Boolean function
$\b(X^{n}):\{0,1\}^{n}\to\{0,1\}$
\begin{equation}
I(\b(X^{n});Y^{n})=H(Y^{n})-H(Y^{n}|\b(X^{n}))\leq1-\binent(\alpha)\label{eq: conjecture inequality}
\end{equation}
where $\binent(\alpha)\dfn-\alpha\log\alpha-(1-\alpha)\log(1-\alpha)$
is the binary entropy function.%
\footnote{Throughout, the logarithm $\log(t)$ is on base $2$, while $\ln(t)$
is the natural logarithm.%
}
\end{conjecture}
Since the \emph{dictator} function $\dic(x^{n})\dfn x_{1}$ (or any
other coordinate) achieves this upper bound with equality, then loosely
stated, Conjecture \ref{conj: Boolean function conjecture} claims
that dictator is the most ``informative'' one-bit quantization of
$X^{n}$ in terms of reducing the entropy of $Y^{n}$. Despite considerable
effort in several directions (e.g. \cite{Boolean_conjecture,anantharam,Chandar_Tcham,Ordentlich_Shayevitz_Weinstein}),
Conjecture \ref{conj: Boolean function conjecture} remains generally
unsettled. Recently, it was shown in \cite{Alex} that Conjecture
\ref{conj: Boolean function conjecture} holds for very noisy channels,
to wit for all $\alpha\geq\frac{1}{2}-\alpha^{*}$, for some absolute
constant $\alpha^{*}>0$. 

From a different perspective, defining $Q_{k}\dfn\P[Y_{k}=1|Y^{k-1},\b(X^{n})]$,
and using the chain rule, we can write 
\begin{align}
H(Y^{n}|\b(X^{n})) & =\sum_{k=1}^{n}H(Y_{k}|Y^{k-1},\b(X^{n}))\nonumber \\
 & =\sum_{k=1}^{n}\E\left[\ell_{\st[log]}(Y_{k},Q_{k})\right]
\end{align}
where $\ell_{\st[log]}(b,q)\dfn-\log[1-q-b(1-2q)]$ is the \emph{binary
logarithmic loss }function\emph{.}%
\footnote{The first argument of $\ell_{\st[log]}(b,q)$ represents the outcome
of the next bit, and the second argument is the probability assignment
for the bit being $1$. %
}\emph{ } Thus, the most informative Boolean function $\b(x^{n})$
can also be interpreted as the one that minimizes the (expected) \emph{sequential
prediction cost} incurred when predicting the sequence $\{Y_{k}\}$
from its past, under logarithmic loss, and given $\b(X^{n})$. It
is important to note that the logarithmic loss function is \textit{proper},
i.e., corresponds to a \emph{proper scoring rule} \cite{gneiting2007strictly}.%
\footnote{Scoring rules are typically defined in the literature as a quantity
to maximize, hence are the negative of cost functions.%
} This means that using the true conditional distribution $Q_{k}$
as the predictor for $Y_{k}$ is guaranteed to minimize the expected
prediction cost at time $k$.

Given the above interpretation, it seems natural to ask the same question
for other loss functions. Namely, what is the minimal sequential prediction
cost of $\{Y_{k}\}$ incurred under a general loss function $\ell:\{0,1\}\times[0,1]\to\mathbb{R}_{+}$,
\[
\l(Y^{n}|\b(X^{n}))\dfn\sum_{k=1}^{n}\E\left[\ell(Y_{k},Q_{k})\right],
\]
and what is the associated optimal Boolean function $\b(x^{n})$?
Specifically, it makes sense to consider proper loss functions, as
for such functions the optimal prediction strategy is ``honest''.
The family of proper loss functions contains many members besides
the logarithmic loss; in fact, the exact characterization of this
family is well known \cite{gneiting2007strictly}. In this work we
focus on another prominent member of this family, the \emph{quadratic
loss function}. This loss function is simply the quadratic distance
between the expected guess and the outcome. In the binary case, it
is given by $\ell_{\st[quad]}(b,q)\dfn(b-q)^{2}$. Following that,
we can define the \emph{sequential mean squared error} (SMSE) to be
the (expected) sequential prediction cost of $Y^{n}$ incurred under
quadratic loss given $\b(X^{n})$, namely 
\begin{align}
\m(Y^{n}|\b(X^{n})) & \dfn\sum_{k=1}^{n}\E\left[\ell_{\st[quad]}(Y_{k},Q_{k})\right]\nonumber \\
 & =\sum_{k=1}^{n}\E\left[Q_{k}(1-Q_{k})\right]\nonumber \\
 & \dfn\sum_{k=1}^{n}\m(Y_{k}|Y^{k-1},\b(X^{n})).\label{eq: SMSE quadratic def}
\end{align}

In what follows, we show that for $\alpha=0$ (noiseless channel)
the SMSE is asymptotically minimized by the majority function.%
\footnote{In fact, for balanced functions, it is trivially maximized by the
dictator.%
} We further show that majority is better than dictator for small $\alpha$.
This might tempt one to conjecture that majority is always asymptotically
optimal for SMSE. However, we show that dictator is in fact better
than majority for $\alpha$ close to $\frac{1}{2}$. Intuitively,
it would seem that dictator is in some sense the function ``least
affected'' by noise, and hence while majority is better at weak noise,
dictator ``catches up'' with it as the noise increases. This intuition
sits well Conjecture \ref{conj: Boolean function conjecture}, since
for logarithmic loss all (balanced) functions are equally good at
$\alpha=0$. We conjecture that the optimal function under quadratic
loss must be close to majority for $\alpha\approx0$, and close to
dictator for $\alpha\approx\frac{1}{2}$. The validity of this conjecture
would imply in particular that, in contrast to the common belief in
the logarithmic loss case, for quadratic loss there is no single sequence
of Boolean functions that is simultaneously (asymptotically) optimal
at all noise levels.

\section{Results}

Let $\Hamw(x_{k}^{m})$ be the Hamming weight of $x_{k}^{m}$. We
denote the majority function by $\maj(x^{n})$, which is equal to
$1$ whenever $\Hamw(x^{n})>\frac{n}{2}$, and $0$ whenever $\Hamw(x^{n})<\frac{n}{2}$.
When $n$ is odd this definition is unambiguous, but when $n$ is
even, the values of $\maj(x^{n})$ when $\Hamw(x^{n})=\frac{n}{2}$
are not defined, and any arbitrary choice of assignment of values
to $\maj(x^{n})$ is proper for our needs.

In the noiseless case ($\alpha=0$), the assertion in Conjecture \ref{conj: Boolean function conjecture}
for the logarithmic loss is trivial, and equality is obtained for
any \emph{balanced} function ($\P[\b(X^{n})=1]=\frac{1}{2}$), and
specifically, for the dictator function. By contrast, for quadratic
loss, finding the optimal function seems far from trivial even for
$\alpha=0$. In the next theorem we provide a lower bound on the noiseless
SMSE for any Boolean function, and show that the majority function
asymptotically achieves it. 
\begin{thm}[Noiseless case]
\label{thm: noiseless}For any Boolean function $\b(X^{n})$ 
\begin{equation}
\m(X^{n}|\b(X^{n}))\geq\frac{n-2\ln2}{4},\label{eq: noiseless lower bound}
\end{equation}
and for majority 
\begin{equation}
\m(X^{n}|\maj(X^{n}))\leq\frac{n-2\ln2}{4}+o(1).\label{eq: noiseless majority}
\end{equation}

\end{thm}
Clearly, for dictator 
\[
\m(X^{n}|\dic(X^{n}))=\frac{n-1}{4}
\]
which is strictly worse than the SMSE of the majority function. In
fact, it is easy to see that dictator in fact maximizes the SMSE.

The minimal SMSE for moderate values of $n$, can be found efficiently.
The idea is to trace, for each $n$, the optimal functions $\{\b_{w}^{(n)}\}_{w\in\{0,1,\ldots,2^{n}\}}$
under a weight constraint
\[
\b_{w}^{(n)}\dfn\argmin_{\b(\cdot):\;\left|\left\{ x^{n}:\;\b(x^{n})=1\right\} \right|=w}\m(X^{n}|\b(X^{n})).
\]
The optimal function $\b^{(n)}$ is then given by optimizing over
$w$, i.e., 
\[
\b^{(n)}\dfn\argmin_{w\in\{0,1,\ldots,2^{n}\}}\m(X^{n}|\b_{w}^{(n)}(X^{n})).
\]
Now, assuming that $\{\b_{w}^{(n)}\}$ were found for all input of
size less than $n$, $\b_{w}^{(n+1)}$ can be found by partitioning
it into two functions of input size $n$ - one pertaining to $x_{1}=0$
and the other to $x_{1}=1$. Indeed, observing (\ref{eq: SMSE quadratic def})
for any given function $\b(\cdot)$, it can be noted that the SMSE
of the first time point, i.e., $\m(X_{k}|X^{k-1},\b(X^{n}))$, depends
only on the weights $w_{0}=\left|\left\{ x_{2}^{n}:\;\b(0,x_{2}^{n})=1\right\} \right|$
and $w_{1}=\left|\left\{ x_{2}^{n}:\;\b(1,x_{2}^{n})=1\right\} \right|$.
Further, for any given $(w_{0},w_{1}):\; w=w_{0}+w_{1}$, the SMSE
of all other time points, i.e. $\sum_{k=2}^{n}\m(X_{k}|X^{k-1},\b(X^{n}))$,
is minimized by setting
\[
\b(0,x_{2}^{n+1})=\b_{w_{0}}^{(n)}(x_{2}^{n+1})
\]
and 
\[
\b(1,x_{2}^{n+1})=\b_{w_{1}}^{(n)}(x_{2}^{n+1}).
\]
Hence, given $\{\b_{w}^{(n)}\}$ for all $n$, we can find $\b_{w}^{(n+1)}$
by simply going over all possible allocation of weights $(w_{0},w_{1}):\; w=w_{0}+w_{1}$.
The output of such an algorithm is shown in Table \ref{tab:SMSE}
for moderate input sizes. It can be seen that majority is optimal
for $n=3$, but not for $n=5,7,9,11$. However, Theorem \ref{thm: noiseless}
states that the difference tends to $0$, as $n\to\infty$. For $n=5$,
the optimal function disagrees with majority on $4$ inputs.

\begin{table}
\begin{centering}
\begin{tabular}{|c|c|c|c|c|}
\hline 
 & $\m(X^{n}|\maj(X^{n}))$ & $\min_{\b(\cdot)}\m(X^{n}|\b(X^{n}))$ & Excess SMSE of majority & Lower bound (\ref{eq: noiseless lower bound})\tabularnewline
\hline 
\hline 
$n=3$ & $0.4792$ & $0.4792$ & $0$ & $0.4034$\tabularnewline
\hline 
$n=5$ & $0.9676$ & $0.9686$ & $0.0010$ & $0.9034$\tabularnewline
\hline 
$n=7$ & $1.4552$ & $1.4618$ & $0.0066$ & $1.4034$\tabularnewline
\hline 
$n=9$ & $1.9483$ & $1.9569$ & $0.0086$ & $1.9034$\tabularnewline
\hline 
$n=11$ & $2.4435$ & $2.4532$ & $0.0097$ & $2.4034$\tabularnewline
\hline 
\end{tabular}\protect\caption{SMSE of majority and SMSE of the optimal function, and (\ref{eq: noiseless lower bound}).
\label{tab:SMSE}}

\par\end{centering}

\end{table}

Next, we consider the noisy case $\alpha\in(0,\frac{1}{2}]$, and
derive a simple lower bound on the noisy SMSE for any Boolean function.
Then, we provide an upper bound and a lower bound for the SMSE of
majority.%
\footnote{Eqs. (\ref{eq: noiseless lower bound}) and (\ref{eq: noiseless majority})
of Theorem \ref{thm: noiseless} can be obtained as special cases
of (\ref{eq: noisy lower bound}) and (\ref{eq: noisy majority upper bound})
of Theorem \ref{thm: noisy}, by setting $\alpha=0$, but since the
proof of the noisy case is based on Theorem \ref{thm: noiseless},
we have separated the results on the noiseless and noisy cases to
two different theorems.%
}
\begin{thm}[Noisy case]
\label{thm: noisy}For any Boolean function $\b(X^{n})$ 
\begin{equation}
\m(Y^{n}|\b(X^{n}))\geq\frac{n-2\ln2\cdot(1-2\alpha)^{2}}{4}.\label{eq: noisy lower bound}
\end{equation}
Furthermore, for majority 
\begin{equation}
\m(Y^{n}|\maj(X^{n}))\leq\frac{n-2\ln2\cdot(1-2\alpha)^{2}\cdot[1-\mu(\alpha)]}{4}+o(1),\label{eq: noisy majority upper bound}
\end{equation}
where 
\begin{equation}
\mu(\alpha)\dfn\binent\left(\frac{\arccos(1-2\alpha)}{\pi}\right),\label{eq: mu definition}
\end{equation}
and 
\begin{equation}
\m(Y^{n}|\maj(X^{n}))\geq\frac{n-\frac{1}{2\pi\alpha(1-\alpha)}(1-2\alpha)^{2}}{4}-O\left(\left(1-2\alpha\right)^{4}\right)+o(1).\label{eq: noisy majority lower bound}
\end{equation}

\end{thm}
Since a straightforward derivation shows that for the dictator function,
\[
\m(Y^{n}|\dic(X^{n}))=\frac{n-(1-2\alpha)^{2}}{4},
\]
the above theorem implies that majority is asymptotically better than
dictator for all\textbf{ }$\alpha\in[0,\underline{\alpha}]$ where
$\underline{\alpha}\approx0.0057$, but that on the other hand, there
exists $\overline{\alpha}<\frac{1}{2}$ such that dictator is better
than majority for all $\alpha\in[\overline{\alpha},\frac{1}{2})$.
\begin{rem}
To improve the SMSE, unbalanced majority functions $\maj_{q}(\cdot)$
may be proposed, which assign $1$ to a set of $q\cdot2^{n}$ vectors
of maximal Hamming weight, $q\in(0,1)$. In the noiseless case, such
functions cannot asymptotically improve the SMSE, since the lower
bound is achieved by ordinary majority functions ($q=\frac{1}{2}$).
Furthermore, it can be shown that they offer no improvement even in
the noisy case. Indeed, the noiseless SMSE of such functions is 
\[
\m(X^{n}|\maj(X^{n}))\leq\frac{n-2\ln2\cdot\binent(q)}{4}+o(1),
\]
which is minimized for $q=\frac{1}{2}$. In addition, the effect of
the noise of the SMSE is related to boundary size between vectors
with $\maj_{q}(x^{n})=1$ and vectors with $\maj_{q}(x^{n})=0$. For
any fixed $q\in(0,1)$, the value of $1$ will be assigned by $\maj_{q}(\cdot)$
to vectors of Hamming weight $\frac{n}{2}-O(n^{\nicefrac{n}{2}+\rho})\leq\frac{n}{2}\leq\frac{n}{2}+O(n^{\nicefrac{n}{2}+\rho})$,
which is asymptotically the same as for ordinary majority with $q=\frac{1}{2}$.
So, the boundary size of $\maj_{q}(\cdot)$ is roughly as the boundary
size of $\maj(\cdot)$, and the effect of the noise on the SMSE is
asymptotically the same for all $q\in(0,1)$. Since the noiseless
SMSE for $q=\frac{1}{2}$ is minimal, this seems to be the optimal
choice even in the presence of noise $(\alpha\in(0,\frac{1}{2})$). 
\end{rem}
The proofs of Theorems \ref{thm: noiseless} and \ref{thm: noisy}
appear in Sections \ref{sec:The-Noiseless-Case} and \ref{sec:The-Noisy-Case},
respectively, and will shortly outlined. Throughout the proofs, we
will only consider positive sequences of $n$ and so Landau notations
should be interpreted with a positive sign. For example, if $a_{n}=\Theta(n)$
then $a_{n}$ is a positive sequence, increasing approximately linearly.
In addition, we will denote the \emph{binary divergence} by $\bindiv(\alpha||\beta)\dfn\alpha\log\frac{\alpha}{\beta}+(1-\alpha)\log\frac{(1-\alpha)}{(1-\beta)}$,
and the support of a random vector $X^{n}$ by ${\cal S}_{X^{n}}\dfn\{x^{n}:\;\P(X^{n}=x^{n})>0\}$.
For brevity, we ignore integer constraints throughout the paper, as
they do no affect the results.

\section{Proof of the Noiseless Case Theorem\label{sec:The-Noiseless-Case}}

In this section, we consider the noiseless case $\alpha=0$, namely
where $X^{n}=Y^{n}$ with probability $1$, and prove Theorem \ref{thm: noiseless}.
The outline of the proof is as follows. To prove the lower bound (\ref{eq: noiseless lower bound})
on the SMSE, we use the binary Pinsker inequality to upper bound the
quadratic loss using the binary divergence. To prove that majority
asymptotically achieves this lower bound, we first note that since
$\maj(X^{n})$ is a balanced function, its value does not help predict
$X_{1}$ at all, and similarly, the gain in SMSE from knowing $\maj(X^{n})$
at the first few time points is negligible. In the same spirit, at
the last time point, the value of $\maj(X^{n})$ is only useful if
$\Hamw(x^{n-1})=\frac{n}{2}$ (assuming odd $n$), which occurs with
negligible probability, and similarly, the gain at the last few time
points due to value of $\maj(X^{n})$ is also negligible. Hence, the
gain in prediction cost from knowing $\maj(X^{n})$ is mainly obtained
in the ``middle'' time points. However, even at those time points,
the gain is moderate and the probability of the next bit, given the
past and $\maj(X^{n})$ is still close to $\frac{1}{2}$, with high
probability. So, as Pinsker's inequality is tight around $\frac{1}{2}$,
the quadratic loss function can be replaced with a function of the
binary divergence. In turn, the binary divergence is related to the
entropy, conditioned on $\maj(X^{n})$. The entropy is simpler to
handle, since conditioned on $\maj(X^{n})$ the reduction in the entropy
of $X^{n}$ is $1$ bit, and this leads directly to (\ref{eq: noiseless majority}).
It should be noted that while the above intuition is fairly simple,
a careful analysis is required for the proof, since a constant deviation
$\frac{2\ln2}{4}$ from $\frac{n}{4}$ is sought, which does not depend
on $n$. We begin with proving the lower bound (\ref{eq: noiseless lower bound})
using Pinsker's inequality.
\begin{IEEEproof}[Proof of (\ref{eq: noiseless lower bound})]
Suppose that $\P[\b(X^{n})=1]=q$, and let $P_{k}\dfn\P[X_{k}=1|X^{k-1},\b(X^{n})=1]$.
Conditioning on $\b(X^{n})=1$, $X^{n}$ is distributed uniformly
over a set of size $q\cdot2^{n}$ and thus 
\begin{align}
\m(X^{n}|\b(X^{n})=1) & =\sum_{k=1}^{n}\E\left[P_{k}(1-P_{k})\right]\nonumber \\
 & =\frac{n}{4}-\sum_{k=1}^{n}\E\left[\left(P_{k}-\frac{1}{2}\right)^{2}\right]\nonumber \\
 & \trre[\geq,a]\frac{n}{4}-\frac{2\ln2}{4}\sum_{k=1}^{n}\E\left[\bindiv(P_{k}||\nicefrac{1}{2})\right]\nonumber \\
 & =\frac{n}{4}-\frac{2\ln2}{4}\sum_{k=1}^{n}\E\left[1-\binent(P_{k})\right]\nonumber \\
 & =\frac{n}{4}-\frac{2\ln2}{4}\left[n-H(X^{n}|\b(X^{n})=1)\right]\nonumber \\
 & =\frac{n}{4}+\frac{2\ln2\log(q)}{4}
\end{align}
where $(a)$ is using a binary version of Pinsker's inequality \cite[p. 370, Eq. (11.139)]{Cover:2006:EIT:1146355}
\begin{equation}
\bindiv(\alpha||\beta)\geq\frac{4}{2\ln2}(\alpha-\beta)^{2}\label{eq: Pinsker inequality binary}
\end{equation}
(where equality is achieved iff $\alpha=\beta$). Deriving a similar
bound for the event $\b(X^{n})=0$, we obtain (\ref{eq: noiseless lower bound})
from 
\begin{align}
\m(X^{n}|\b(X^{n})) & =q\cdot\m(X^{n}|\b(X^{n})=1)+(1-q)\cdot\m(X^{n}|\b(X^{n})=0)\nonumber \\
 & \geq\frac{n}{4}-\frac{2\ln2\cdot\binent(q)}{4}\nonumber \\
 & \geq\frac{n}{4}-\frac{2\ln2}{4}.\label{eq: averaging over b}
\end{align}

\end{IEEEproof}
Proving the asymptotic achievability of the lower bound (\ref{eq: noiseless lower bound})
by the majority function is more intricate, and is based on the asymptotic
achievability of equality in Pinsker's inequality (\ref{eq: Pinsker inequality binary}).
We will need several definitions and lemmas. 
\begin{defn}
A vector $v^{n}\in\{0,1\}^{n}$ is termed \emph{$t$-majority vector}
if $\Hamw(v^{n})\geq tn$, where $t\in[0,1]$ is referred to as the
\emph{threshold}. A random vector $V^{n}$ will be termed\emph{ $t$-majority
random vector} if it is uniformly distributed over all \emph{$t$-}majority
vectors of length $n$\emph{.} Let $\zeta_{n}(t)$ be the minimal
integer larger or equal to $tn$. A random vector $V^{n}$ will be
termed\emph{ pseudo $t$-majority random vector} if it is uniformly
distributed over all \emph{$t$-}majority vectors of length $n$\emph{,
}except possibly for some set ${\cal D}_{n}$, such that $\Hamw(v^{n})=\zeta_{n}(t)$
for all $v^{n}\in{\cal D}^{n}$, and there exists $v^{n}\in{\cal S}_{V^{n}}$
such that $\Hamw(v^{n})=\zeta_{n}(t)$. For brevity, we will sometime
omit the parameter $t$ when $t=\frac{1}{2}$.

The first lemma provides an approximation for the marginal distributions
of a $t$-majority random vector. \end{defn}
\begin{lem}
\label{lem:marginal of t-majority}Let $\eta\in[0,\frac{1}{2})$ be
given. Then, if $V^{n}$ is a\textcolor{red}{{} }pseudo $t$-majority
random vector, 
\[
\max\left[\frac{1}{2},t\right]\leq\P[V_{k}=1]\leq\max\left[\frac{1}{2},t\right]+O_{\eta}\left(\frac{1}{n^{\nicefrac{1}{2}-\eta}}\right)
\]
for all $k\in[n]$. \end{lem}
\begin{IEEEproof}
See Appendix \ref{sec:Miscellaneous-Proofs}.
\end{IEEEproof}
Before we continue, we shortly comment on notation conventions. There
is obviously a difference between a majority random vector of length
$k$, and the first $k$ coordinates of a majority random vector of
length $n$, when $k<n$. Nonetheless, to avoid double indexing, we
will assume that $n$ is large enough but fixed, and the indices of
$V^{n}$ will denote the corresponding components, e.g. $V_{k}^{k+m}$
are the components $(V_{k},\ldots,V_{k+m})$ of the majority random
vector $V^{n}$.

The following lemma shows that if $m_{n}$ increases slowly enough,
then the entropy loss of $1$ bit of a majority random vector $V^{n}$,
compared to the entropy of a uniform binary i.i.d. random vector,
is mainly due to the entropy of the middle part of the vector $V_{m_{n}}^{n-m_{n}}$.
In other words, the conditional entropies of the beginning and end
parts are close to their maximal values, given by their length. 
\begin{lem}
\label{lem: entropy of partial vector}Let $\rho\in(0,\frac{1}{4})$
and $m_{n}=O(n^{\nicefrac{1}{4}-\rho})$. Then, for a majority random
vector $V^{n}$\textbf{ 
\[
H(V_{m_{n}+1}^{n-m_{n}})\leq n-1-2m_{n}+o(1).
\]
}\end{lem}
\begin{IEEEproof}
See Appendix \ref{sec:Miscellaneous-Proofs}.
\end{IEEEproof}

The following corollary is a weakening of lemma \ref{lem: entropy of partial vector}. 
\begin{cor}
\label{cor: entropy of partial vector}Let $\rho\in(0,\frac{1}{4})$
and $m_{n}=O(n^{\nicefrac{1}{4}-\rho})$. Then, for a majority random
vector $V^{n}$\textbf{ 
\[
H(V_{1}^{n-m_{n}})\leq n-1-m_{n}+o(1).
\]
}
\end{cor}
Now, consider a time index $k$ which is sufficiently far from the
last index $n$. In the next lemma, we bound the probability that
at time $k$, the number of ones in the vector is still significantly
less than the minimal weight $\frac{k}{2}$ of vectors in the support
of a majority random vector of length $k$.
\begin{lem}
\label{lem: exp small probability}Let $m_{n}$ be an increasing positive
sequence, and let $\rho\in(0,1)$ be given. Then, for all majority
random vectors $V^{n}$ with sufficiently large $n$,
\[
\P\left[\Hamw(V_{1}^{k})\leq\frac{k-1}{2}-(n-k+1)^{\nicefrac{1}{2}+\rho}\right]\leq2^{-\Omega(m_{n}^{2\rho})},
\]
for all $k\in[n-m_{n}]$.\end{lem}
\begin{IEEEproof}
See Appendix \ref{sec:Miscellaneous-Proofs}.
\end{IEEEproof}
We are now ready to prove that majority functions are asymptotically
optimal. 
\begin{IEEEproof}[Proof of (\ref{eq: noiseless majority})]
 Let $\rho\in(0,\nicefrac{1}{8})$ be given, and define $m_{n}\dfn n^{\nicefrac{1}{4}-\rho}$.
Let us define $V^{n}$ as the random vector distributed as $X^{n}$
conditioned on $\maj(X^{n})=1$. Clearly, $V^{n}$ is a majority random
vector. For any given $k\in[n-m_{n}]$ let us define the events 
\begin{align}
{\cal A}_{k} & \dfn\left\{ \Hamw(V_{1}^{k})\geq\frac{k-1}{2}-(n-k+1)^{\nicefrac{1}{2}+\rho}\right\} \nonumber \\
 & =\left\{ \Hamw(V_{1}^{k})\geq\frac{n}{2}-r_{k}+1\right\} \label{eq: large probability set}
\end{align}
where $r_{k}\dfn\frac{(n-k+1)}{2}+(n-k+1)^{\nicefrac{1}{2}+\rho}$.
Now, letting $P_{k}\dfn\P[V_{k}=1|V^{k-1}]$ we have 
\begin{align}
\m(X^{n}|\maj(X^{n})=1) & =\sum_{k=1}^{n}\E\left[P_{k}(1-P_{k})\right]\nonumber \\
 & =\frac{n}{4}-\sum_{k=1}^{n}\E\left[\left(P_{k}-\frac{1}{2}\right)^{2}\right]\nonumber \\
 & \leq\frac{n}{4}-\sum_{k=1}^{n-m_{n}}\E\left[\left(P_{k}-\frac{1}{2}\right)^{2}\right]\nonumber \\
 & \leq\frac{n}{4}-\sum_{k=1}^{n-m_{n}}\sum_{v^{k-1}\in{\cal A}_{k-1}}\P\left[V^{k-1}=v^{k-1}\right]\E\left[\left(P_{k}-\frac{1}{2}\right)^{2}\vert V^{k-1}=v^{k-1}\right].
\end{align}
Now, let $v^{k-1}\in{\cal A}_{k-1}$. Conditioning on $V^{k-1}=v^{k-1}$,
we have that $V_{k}^{n}$ is a $t$-majority random vector of length
$n-k+1\geq m_{n}$, and its threshold $t$ is less than 
\begin{align}
t\leq\frac{r_{k}}{n-k+1} & =\frac{1}{2}+\frac{1}{(n-k+1)^{\nicefrac{1}{2}-\rho}}\nonumber \\
 & \leq\frac{1}{2}+\frac{1}{m_{n}^{\nicefrac{1}{2}-\rho}}.
\end{align}
So, assuming that $n$ is large enough, Lemma \ref{lem:marginal of t-majority}
(with $\eta<\rho$) implies that conditioned on the event $V^{k-1}=v^{k-1}$
with $v^{k-1}\in{\cal A}_{k}$ 
\[
\frac{1}{2}\leq P_{k}\leq\frac{1}{2}+\frac{1}{m_{n}^{\nicefrac{1}{2}-\rho}}+\frac{1}{m_{n}^{\nicefrac{1}{2}-\eta}}\leq\frac{1}{2}+O_{\eta}\left(\frac{1}{n^{\nicefrac{1}{8}-\rho}}\right),
\]
for all $k\in[n-m_{n}]$. Consequently, as Pinsker's inequality is
tight around $\frac{1}{2}$, 
\[
\left(P_{k}-\frac{1}{2}\right)^{2}\geq\left[1-o(1)\right]\frac{\ln2}{2}\bindiv(P_{k}||\nicefrac{1}{2})
\]
and so 
\begin{align}
\m(X^{n}|\maj(X^{n})=1) & \leq\frac{n}{4}-\frac{2\ln2}{4}\left[1-o(1)\right]\times\nonumber \\
 & \hphantom{\leq}\sum_{k=1}^{n-m_{n}}\sum_{v^{k-1}\in{\cal A}_{k-1}}\P\left[V^{k-1}=v^{k-1}\right]\E\left[\bindiv(P_{k}||\nicefrac{1}{2})\vert V^{k-1}=v^{k-1}\right].
\end{align}
Denoting $\tau_{k}\dfn\P\left[V^{k}\not\in{\cal A}_{k}\right]$, we
have
\begin{align}
\E\left[\bindiv(P_{k}||\nicefrac{1}{2})\right] & =\sum_{v^{k-1}\in{\cal A}_{k-1}}\P\left[V^{k-1}=v^{k-1}\right]\E\left[\bindiv(P_{k}||\nicefrac{1}{2})\vert V^{k-1}=v^{k-1}\right]\nonumber \\
 & \hphantom{=}+\sum_{v^{k-1}\not\in{\cal A}_{k-1}}\P\left[V^{k-1}=v^{k-1}\right]\E\left[\bindiv(P_{k}||\nicefrac{1}{2})\vert V^{k-1}=v^{k-1}\right]\nonumber \\
 & \leq\sum_{v^{k-1}\in{\cal A}_{k-1}}\P\left[V^{k-1}=v^{k-1}\right]\E\left[\bindiv(P_{k}||\nicefrac{1}{2})\vert V^{k-1}=v^{k-1}\right]+\tau_{k-1},\label{eq: conditional and unconditional divergence}
\end{align}
because $\bindiv(P_{k}||\nicefrac{1}{2})=1-\binent(P_{k})\leq1$.
Hence, 
\begin{align}
\m(X^{n}|\maj(X^{n})=1) & \leq\frac{n}{4}-\frac{2\ln2}{4}\left[1-o(1)\right]\sum_{k=1}^{n-m_{n}}\left\{ \E\left[\bindiv(P_{k}||\nicefrac{1}{2})\right]-\tau_{k-1}\right\} \nonumber \\
 & \trre[\leq,a]\frac{n}{4}-\frac{2\ln2}{4}\left[1-o(1)\right]\left[n-m_{n}-H(V_{1}^{n-m_{n}})\right]\nonumber \\
 & \hphantom{=}+\left[1-o(1)\right]\sum_{k=1}^{n-m_{n}}2^{-cm_{n}^{2\rho}}\nonumber \\
 & \trre[\leq,b]\frac{n}{4}-\left[1-o(1)\right]\frac{2\ln2}{4}+o(1)+n2^{-cm_{n}^{2\rho}}\nonumber \\
 & =\frac{n}{4}-\frac{2\ln2}{4}+o(1),\label{eq: last bound on MMSE for majority}
\end{align}
where $(a)$ is using the chain rule, $\bindiv(P_{k}||\nicefrac{1}{2})=1-\binent(P_{k})$,
and since from Lemma \ref{lem: exp small probability}, for some $c>0$
we have $\tau_{k}\leq2^{-cm_{n}^{2\rho}}$ for all $k\in[n-m_{n}]$,
and $(b)$ is using Corollary \ref{cor: entropy of partial vector}. 

Finally, from symmetry, conditioning on $\maj(X^{n})=0$ we have
\[
\m(X^{n}|\maj(X^{n})=0)\leq\frac{n}{4}-\frac{2\ln2}{4}+o(1)
\]
and so (\ref{eq: noiseless majority}) is obtained by averaging over
$\maj(X^{n})$ (as in (\ref{eq: averaging over b})).
\end{IEEEproof}

\section{Proof of the Noisy Case Theorem \label{sec:The-Noisy-Case}}

In this section, we consider the noisy case, and prove Theorem \ref{thm: noisy}.
The outline of the proof is as follows. The lower bound of (\ref{eq: noisy lower bound})
is based on the the result of the noiseless case (\ref{eq: noiseless lower bound}),
while taking into account that a noisy bit $Y_{k}$ is to be predicted
rather than $X_{k}$. To prove (\ref{eq: noisy majority upper bound})
we use the noiseless SMSE of majority (\ref{eq: noiseless majority}),
and quantify the loss in the SMSE conditioned on majority, due to
the fact that noisy past bits $Y^{k-1}$ are observed, rather than
the noiseless $X^{k-1}$. As in the noiseless case, the ``middle''
time points contain most of the loss. In addition, we use a bound
on $H(Y^{n}|\maj(X^{n}))$ based on the \emph{stability} of majority.
Finally, to prove (\ref{eq: noisy majority lower bound}) we use a
different asymptotic lower bound on $H(\maj(X^{n})|Y^{n})$, which
is based on the Gaussian approximation of a binomial random variable,
resulting from the Berry-Essen central limit theorem. We then apply
Pinsker's inequality, as in the noiseless case, to bound the SMSE
via that entropy. 

To prove (\ref{eq: noisy lower bound}) begin with the next lemma,
which states a bound on SMSE of a channel output in terms of the input's
SMSE, for any input distribution.
\begin{lem}
\label{lem: single sample mmse over a channel}For $V\sim\mbox{Bern}(\beta),$
$Z\sim\mbox{Bern}(\alpha)$ independent of $V$, and $W=V+Z$ (modulo-$2$
sum), 
\[
\m(W)=\alpha(1-\alpha)+(1-2\alpha)^{2}\cdot\m(V).
\]
\end{lem}
\begin{IEEEproof}
See Appendix \ref{sec:Miscellaneous-Proofs}.\end{IEEEproof}
\begin{lem}
\label{lem: mmse over a channel}Let $V^{n}\in\{0,1\}^{n}$ be a random
vector, and $W^{n}$ be the output of a BSC with crossover $\alpha$
fed by $V^{n}$, i.e. $W^{n}=V^{n}+Z^{n}$, where $Z^{n}\sim\mbox{Bern}(\alpha)$,
independent of $V^{n}$. Then, 
\[
\m(W^{n})\geq\alpha(1-\alpha)\cdot n+(1-2\alpha)^{2}\cdot\m(V^{n})
\]
with equality if $V^{n}$ is a memoryless random vector.\end{lem}
\begin{IEEEproof}
See Appendix \ref{sec:Miscellaneous-Proofs}.
\end{IEEEproof}
Using the above, we can prove (\ref{eq: noisy lower bound}). 
\begin{IEEEproof}[Proof of (\ref{eq: noisy lower bound})]
Consider any Boolean function $\b(X^{n})$ and suppose that $\P\left[\b(X^{n})=1\right]=q$.
Then, 
\begin{align}
\m(Y^{n}|\b(X^{n})) & \trre[\geq,a]\alpha(1-\alpha)\cdot n+(1-2\alpha)^{2}\cdot\m(X^{n}|\b(X^{n}))\nonumber \\
 & =\alpha(1-\alpha)\cdot n+q(1-2\alpha)^{2}\cdot\m(X^{n}|\b(X^{n})=1)+(1-q)(1-2\alpha)^{2}\cdot\m(X^{n}|\b(X^{n})=0)\nonumber \\
 & \trre[\geq,b]\alpha(1-\alpha)\cdot n+(1-2\alpha)^{2}\cdot\frac{(n-2\ln2)}{4}\nonumber \\
 & \geq\frac{n-(1-2\alpha)^{2}\cdot2\ln2}{4},\label{eq: lower bound on MMSE conditional}
\end{align}
where $(a)$ follows from Lemma \ref{lem: mmse over a channel}, and
$(b)$ follows from (\ref{eq: noiseless lower bound}).
\end{IEEEproof}

To prove (\ref{eq: noisy majority upper bound}), we analyze, in the
next two lemmas, the SMSE of a majority random vector $V^{n}$, and
show that the quadratic loss in the beginning and end of the vector
is close to its maximal value of $\frac{1}{4}$ per bit.
\begin{lem}
\label{lem: MMSE of beginning of vector}Let $m_{n}=O(n^{1-\rho})$
for some $\rho\in(0,1)$. Then, for a majority random vector $V^{n}$\textbf{
\[
\m(V_{1}^{m_{n}})=\sum_{k=1}^{m_{n}}\m(V_{k}|V_{1}^{k-1})\geq\frac{m_{n}}{4}-o(1).
\]
}\end{lem}
\begin{IEEEproof}
See Appendix \ref{sec:Miscellaneous-Proofs}.\end{IEEEproof}
\begin{lem}
\label{lem: MMSE of end of vector}Let $\rho\in(0,\frac{1}{8})$ and
$m_{n}=O(n^{\nicefrac{1}{4}-\rho})$. Then, for a majority random
vector $V^{n}$\textbf{ 
\[
\sum_{k=n-m_{n}+1}^{n}\m(V_{k}|V_{1}^{k-1})\geq\frac{m_{n}}{4}-o(1).
\]
}\end{lem}
\begin{IEEEproof}
See Appendix \ref{sec:Miscellaneous-Proofs}.
\end{IEEEproof}
We also need the following bound on the conditional entropy of the
output, given a value of the majority of the input.
\begin{lem}
\label{lem: output entropy for majority}Let $\mu(\cdot)$ be as defined
in (\ref{eq: mu definition}). Then, 
\[
H(Y^{n}|\maj(X^{n})=1)\leq n-1+\mu(\alpha)+o(1).
\]
\end{lem}
\begin{IEEEproof}
See Appendix \ref{sec:Miscellaneous-Proofs}.
\end{IEEEproof}
We can now prove (\ref{eq: noisy majority upper bound}). 
\begin{IEEEproof}[Proof of (\ref{eq: noisy majority upper bound})]
In (\ref{eq: lower bound on MMSE conditional}), it may be observed
that due to (\ref{eq: noiseless majority}), inequality $(b)$ is
in fact an asymptotic equality, up to an $o(1)$ term. So, it remains
to bound the loss in the inequality $(a)$ of (\ref{eq: lower bound on MMSE conditional}),
which we denote by $\Phi$. Let us also denote $m_{n}=n^{\nicefrac{1}{4}-\rho}$
for some given $\rho\in(0,\frac{1}{4})$. Then, due to symmetry of
the majority function, we may condition on the event $\maj(X^{n})=1$,
and the loss of inequality $(a)$ of (\ref{eq: lower bound on MMSE conditional})
is 
\begin{align}
\Phi & \dfn\m(Y^{n}|\maj(X^{n})=1)-\alpha(1-\alpha)\cdot n-(1-2\alpha)^{2}\cdot\m(X^{n}|\maj(X^{n})=1)\nonumber \\
 & =\sum_{k=1}^{n}\m(Y_{k}|Y^{k-1},\maj(X^{n})=1)-\alpha(1-\alpha)\cdot n-(1-2\alpha)^{2}\cdot\sum_{k=1}^{n}\m(X_{k}|X^{k-1},\maj(X^{n})=1)\nonumber \\
 & \trre[=,a](1-2\alpha)^{2}\cdot\left\{ \sum_{k=1}^{n}\m(X_{k}|Y^{k-1},\maj(X^{n})=1)-\m(X_{k}|X^{k-1},\maj(X^{n})=1)\right\} ,\label{eq: Phi loss}
\end{align}
where $(a)$ is using a derivation similar to (\ref{eq: mmse one sample}). 

First, using Lemma \ref{lem: MMSE of beginning of vector}
\begin{align}
 & \hphantom{\leq}\sum_{k=1}^{m_{n}}\m(X_{k}|Y^{k-1},\maj(X^{n})=1)-\m(X_{k}|X^{k-1},\maj(X^{n})=1)\nonumber \\
 & \leq\frac{m_{n}}{4}-\sum_{k=1}^{m_{n}}\m(X_{k}|X^{k-1},\maj(X^{n})=1)\nonumber \\
 & \leq o(1),\label{eq: Phi loss start of vector}
\end{align}
and similarly, using Lemma \ref{lem: MMSE of end of vector}

\begin{align}
 & \hphantom{\leq}\sum_{k=m_{n}+1}^{n}\m(X_{k}|Y^{k-1},\maj(X^{n})=1)-\m(X_{k}|X^{k-1},\maj(X^{n})=1)\nonumber \\
 & \leq\frac{m_{n}}{4}-\sum_{k=m_{n}+1}^{n}\m(X_{k}|X^{k-1},\maj(X^{n})=1)\nonumber \\
 & \leq o(1).\label{eq: Phi loss end of vector}
\end{align}
Then, from (\ref{eq: noiseless lower bound}) of Theorem \ref{thm: noiseless},
and the symmetry of conditioning $\maj(X^{n})=0$ and $\maj(X^{n})=1$,
we have
\[
\sum_{k=1}^{n}\m(X_{k}|X^{k-1},\maj(X^{n})=1)\geq\frac{n-2\ln(2)}{4},
\]
and
\begin{align}
 & \hphantom{=}\sum_{k=m_{n}+1}^{n-m_{n}}\m(X_{k}|X^{k-1},\maj(X^{n})=1)\nonumber \\
 & =\sum_{k=1}^{n}\m(X_{k}|X^{k-1},\maj(X^{n})=1)-\sum_{k=1}^{m_{n}}\m(X_{k}|X^{k-1},\maj(X^{n})=1)\nonumber \\
 & \hphantom{=}-\sum_{k=n-m_{n}+1}^{n}\m(X_{k}|X^{k-1},\maj(X^{n})=1)\nonumber \\
 & \geq\sum_{k=1}^{n}\m(X_{k}|X^{k-1},\maj(X^{n})=1)-\frac{m_{n}}{4}-\frac{m_{n}}{4}\nonumber \\
 & \geq\frac{n-2m_{n}-2\ln(2)}{4}.\label{eq: MMSE loss middle of vector}
\end{align}
So it remains to upper bound the first term in the sum of (\ref{eq: Phi loss}),
viz. 
\[
\sum_{k=m_{n}+1}^{n-m_{n}}\m(X_{k}|Y^{k-1},\maj(X^{n})=1).
\]
We follow the outline of the proof of (\ref{eq: noiseless majority})
from Theorem \ref{thm: noiseless}. Let us denote the random variables
$P_{k}(X^{k-1})\dfn\P(X_{k}=1|X^{k-1},\maj(X^{n})=1)$, and $R_{k}(Y^{k-1})\dfn\P(X_{k}=1|Y^{k-1},\maj(X^{n})=1)$,
where their arguments will be sometimes omitted for brevity. In what
follows, we will prove the existence of sets ${\cal B}_{k}\subset\{0,1\}^{k}$
such that $\upsilon_{k}\dfn\P\left[Y^{k}\not\in{\cal B}_{k}\right]\leq2^{-\frac{c}{2}m_{n}^{2\rho}}$
for some $c>0$ and for all $k\in\{m_{n}+1,\ldots,n-m_{n}\}$, and
\[
\frac{1}{2}\leq R_{k}(y^{k-1})\leq\frac{1}{2}+O_{\eta}\left(\frac{1}{n^{\nicefrac{1}{8}-\rho}}\right)
\]
for all $y^{k-1}\in{\cal B}_{k-1}$. For $y^{k-1}\in{\cal B}_{k-1}$
Pinsker's inequality is tight and so 
\[
\left(R_{k}(y^{k-1})-\frac{1}{2}\right)^{2}\geq\left[1-o(1)\right]\frac{\ln2}{2}\bindiv(R_{k}(y^{k-1})||\nicefrac{1}{2}).
\]
Hence, 
\begin{align}
 & \hphantom{=}\sum_{k=m_{n}+1}^{n-m_{n}}\m(X_{k}|Y^{k-1},\maj(X^{n})=1)\nonumber \\
 & =\sum_{k=m_{n}+1}^{n-m_{n}}\E\left[R_{k}(1-R_{k})\right]\nonumber \\
 & =\frac{n-2m_{n}}{4}-\sum_{k=m_{n}+1}^{n-m_{n}}\E\left[\left(R_{k}-\frac{1}{2}\right)^{2}\right]\nonumber \\
 & \leq\frac{n-2m_{n}}{4}-\sum_{k=m_{n}+1}^{n-m_{n}}\sum_{y_{1}^{k-1}\in{\cal B}_{k-1}}\P\left[Y^{k-1}=y_{1}^{k-1}\right]\E\left[\left(R_{k}-\frac{1}{2}\right)^{2}|Y^{k-1}=y_{1}^{k-1}\right]\nonumber \\
 & \leq\frac{n-2m_{n}}{4}-\frac{2\ln(2)}{4}\left[1-o(1)\right]\sum_{k=m_{n}+1}^{n-m_{n}}\sum_{y_{1}^{k-1}\in{\cal B}_{k-1}}\P\left[Y^{k-1}=y_{1}^{k-1}\right]\E\left[\bindiv(R_{k}||\nicefrac{1}{2})|Y^{k-1}=y_{1}^{k-1}\right]\nonumber \\
 & \trre[\leq,a]\frac{n-2m_{n}}{4}-\frac{2\ln(2)}{4}\left[1-o(1)\right]\sum_{k=m_{n}+1}^{n-m_{n}}\left\{ \E\left[\bindiv(R_{k}||\nicefrac{1}{2})\right]-\upsilon_{k}\right\} \nonumber \\
 & \trre[\leq,b]\frac{n-2m_{n}}{4}-\frac{2\ln(2)}{4}\left[1-o(1)\right]\sum_{k=m_{n}+1}^{n-m_{n}}\E\left[\bindiv(R_{k}||\nicefrac{1}{2})\right]+o(1)\nonumber \\
 & =\frac{n-2m_{n}}{4}-\frac{2\ln(2)}{4}\left[1-o(1)\right]\left[n-2m_{n}-\sum_{k=m_{n}+1}^{n-m_{n}}H(X_{k}|Y^{k-1},\maj(X^{n})=1)\right]+o(1)\nonumber \\
 & \trre[\leq,c]\frac{n-2m_{n}}{4}-\frac{2\ln(2)}{4}\left[1-o(1)\right]\left[n-2m_{n}-\sum_{k=m_{n}+1}^{n-m_{n}}H(Y_{k}|Y^{k-1},\maj(X^{n})=1)\right]+o(1)\nonumber \\
 & =\frac{n-2m_{n}}{4}-\frac{2\ln(2)}{4}\left[1-o(1)\right]\left[n-2m_{n}-H(Y_{m_{n}+1}^{n-m_{n}}|Y^{m_{n}},\maj(X^{n})=1)\right]+o(1)\nonumber \\
 & \trre[\leq,d]\frac{n-2m_{n}}{4}-\frac{2\ln(2)}{4}\left[1-o(1)\right]\left[n-H(Y^{n}|\maj(X^{n})=1)\right]+o(1)\nonumber \\
 & \trre[\leq,e]\frac{n-2m_{n}}{4}-\frac{2\ln(2)}{4}\left[1+\mu(\alpha)\right]+o(1),\label{eq: Phi loss Y given X middle}
\end{align}
$(a)$ is since, just as in (\ref{eq: conditional and unconditional divergence}),
\[
\E\left[\bindiv(R_{k}||\nicefrac{1}{2})\right]\leq\sum_{y_{1}^{k-1}\in{\cal B}_{k-1}}\P\left[Y^{k-1}=y_{1}^{k-1}\right]\E\left[\bindiv(R_{k}||\nicefrac{1}{2})|Y^{k-1}=y_{1}^{k-1}\right]+\upsilon_{k},
\]
$(b)$ is since $\upsilon_{k}\leq2^{-\frac{c}{2}m_{n}^{2\rho}}$,
$(c)$ is using 
\begin{align}
H(Y_{k}|Y^{k-1},\maj(X^{n})=1) & =H(X_{k}+Z_{k}|Y^{k-1},\maj(X^{n})=1)\nonumber \\
 & \geq H(X_{k}+Z_{k}|Y^{k-1},Z_{k},\maj(X^{n})=1)\nonumber \\
 & =H(X_{k}|Y^{k-1},Z_{k},\maj(X^{n})=1)\nonumber \\
 & =H(X_{k}|Y^{k-1},\maj(X^{n})=1),
\end{align}
where the last equality is since $Z_{k}$ is independent of $(X_{k},Y^{k-1})$.
Transition $(d)$ in (\ref{eq: Phi loss Y given X middle}) follows
from
\begin{align}
H(Y_{n-m_{m}+1}^{n}|Y_{1}^{n-m_{n}},\maj(X^{n})=1) & \trre[\geq,i]H(Y_{n-m_{m}+1}^{n}|X_{1}^{n-m_{n}},\maj(X^{n})=1)\nonumber \\
 & =H(X_{n-m_{m}+1}^{n}+Z_{n-m_{m}+1}^{n}|X_{1}^{n-m_{n}},\maj(X^{n})=1)\nonumber \\
 & \geq H(X_{n-m_{m}+1}^{n}+Z_{n-m_{m}+1}^{n}|X_{1}^{n-m_{n}},Z_{n-m_{m}+1}^{n},\maj(X^{n})=1)\nonumber \\
 & =H(X_{n-m_{m}+1}^{n}|X_{1}^{n-m_{n}},\maj(X^{n})=1)\nonumber \\
 & \trre[\geq,ii]m_{n}-o(1),
\end{align}
where here $(i)$ follows from the data processing theorem and the
fact that $Y_{1}^{n-m_{n}}-X_{1}^{n-m_{n}}-Y_{n-m_{n}+1}^{n}$, and
$(ii)$ follows from (\ref{eq: bound on conditional entropy}) (proof
of Lemma \ref{lem: entropy of partial vector}), and using a similar
bound to $H(Y_{1}^{m_{n}}|Y_{m_{n}+1}^{n},\maj(X^{n})=1).$ Transition
$(e)$ in (\ref{eq: Phi loss Y given X middle}) follows from Lemma
\ref{lem: output entropy for majority}. To conclude, combining (\ref{eq: Phi loss}),(\ref{eq: Phi loss start of vector}),
(\ref{eq: Phi loss end of vector}), (\ref{eq: MMSE loss middle of vector})
and (\ref{eq: Phi loss Y given X middle}) implies that 
\[
\Phi\leq(1-2\alpha)^{2}\cdot\frac{2\ln2}{4}\mu(\alpha)+o(1),
\]
which, together with (\ref{eq: lower bound on MMSE conditional})
implies (\ref{eq: noisy majority upper bound}). 

To complete the proof, it remains to assert the existence of the sets
${\cal B}_{k}$. To this end, recall that in the proof of (\ref{eq: noiseless majority})
in Section \ref{sec:The-Noiseless-Case}, we have defined the sets

\[
{\cal A}_{k}\dfn\left\{ \Hamw(V_{1}^{k})\geq\frac{k-1}{2}-(n-k+1)^{\nicefrac{1}{2}+\rho}\right\} 
\]
(cf. (\ref{eq: large probability set})) and showed that $\frac{1}{2}\leq P_{k}(x^{k-1})\leq\frac{1}{2}+O\left(\nicefrac{1}{n^{\nicefrac{1}{8}-\rho}}\right)$
for all $x^{k-1}\in{\cal A}_{k-1}$. In addition, Lemma \ref{lem: exp small probability}
implied that there that there exists $c>0$ such that $\P\left[X^{k}\not\in{\cal A}_{k}\right]\leq2^{-cm_{n}^{2\rho}}$
for all $k\in\{m_{n}+1,\ldots,n-m_{n}\}$. Now, note that 
\begin{align}
R_{k}(Y^{k-1}) & =\P(X_{k}=1|Y^{k-1},\maj(X^{n})=1)\nonumber \\
 & =\sum_{x^{k-1}}\P\left(X^{k-1}=x^{k-1}|Y^{k-1},\maj(X^{n})=1\right)\cdot\P\left(X_{k}=1|X^{k-1}=x^{k-1},Y^{k-1},\maj(X^{n})=1\right)\nonumber \\
 & =\sum_{x^{k-1}}\P\left(X^{k-1}=x^{k-1}|Y^{k-1},\maj(X^{n})=1\right)\cdot P_{k}(x^{k-1}),
\end{align}
so $R_{k}(Y^{k-1})$ is just an averaging of $P_{k}(x^{k-1})$. Since
$P_{k}(x^{k-1})\geq\frac{1}{2}$ for all $x^{k-1}$, this immediately
implies $R_{k}(y^{k-1})\geq\frac{1}{2}$. On the other hand
\begin{align}
R_{k}(Y^{k-1}) & =\sum_{x^{k-1}\in{\cal A}_{k-1}}\P\left(X^{k-1}=x^{k-1}|Y^{k-1},\maj(X^{n})=1\right)\cdot P_{k}(x^{k-1})\nonumber \\
 & \hphantom{=}+\sum_{x^{k-1}\not\in{\cal A}_{k-1}}\P\left(X^{k-1}=x^{k-1}|Y^{k-1},\maj(X^{n})=1\right)\cdot P_{k}(x^{k-1})\nonumber \\
 & \leq\frac{1}{2}+O\left(\frac{1}{n^{\nicefrac{1}{8}-\rho}}\right)+\P\left(X^{k-1}\not\in{\cal A}_{k-1}|Y^{k-1},\maj(X^{n})=1\right),
\end{align}
where we have bounded the first term using $P_{k}(x^{k-1})\leq\frac{1}{2}+O\left(\nicefrac{1}{n^{\nicefrac{1}{8}-\rho}}\right)$
for all $x^{k-1}\in{\cal A}_{k-1}$, and we have bounded the second
term simply by using $P_{k}(x^{k-1})\leq1$. Let us inspect the random
variable $\P[X^{k-1}\not\in{\cal A}_{k-1}|Y^{k-1},\maj(X^{n})=1]$.
We know that its expected value satisfies
\[
\E\left[\P\left(X^{k-1}\not\in{\cal A}_{k-1}|Y^{k-1},\maj(X^{n})=1\right)\right]=\P\left(X^{k-1}\not\in{\cal A}_{k-1}|\maj(X^{n})=1\right)\leq2^{-cm_{n}^{2\rho}}.
\]
So, for any given $\eta>0$ Markov's inequality implies that
\[
\P\left[\P\left(X^{k-1}\not\in{\cal A}_{k-1}|Y^{k-1},\maj(X^{n})=1\right)\geq2^{\eta m_{n}^{2\rho}}2^{-cm_{n}^{2\rho}}\right]\leq2^{-\eta m_{n}^{2\rho}}.
\]
Choosing, e.g., $\eta=\frac{c}{2}$ we get that there exists a set
${\cal B}_{k}$ whose probability is larger than $1-2^{-\frac{c}{2}m_{n}^{2\rho}}$
such that 
\[
\P\left(X^{k-1}\not\in{\cal A}_{k-1}|Y^{k-1},\maj(X^{n})=1\right)\leq2^{-\frac{c}{2}m_{n}^{2\rho}}
\]
for all $y^{k-1}\in{\cal B}_{k}$. For this set, we have
\[
R_{k}(Y^{k-1})\leq\frac{1}{2}+O\left(\frac{1}{n^{\nicefrac{1}{8}-\rho}}\right)+2^{-\frac{c}{2}m_{n}^{2\rho}}=\frac{1}{2}+O\left(\frac{1}{n^{\nicefrac{1}{8}-\rho}}\right),
\]
as required. 
\end{IEEEproof}
To prove (\ref{eq: noisy majority lower bound}) we first need the
following approximation to the entropy of majority functions. 
\begin{lem}[\cite{Or_private}]
\label{lem: exact entropy of majority}We have
\begin{equation}
H(\maj(X^{n})|Y^{n})=\E\left\{ \binent\left[Q\left(\frac{\left|G(1-2\alpha)\right|}{\sqrt{4\alpha(1-\alpha)}}\right)\right]\right\} +o(1)\label{eq: Gaussian approximation}
\end{equation}
where $G\sim{\cal N}(0,1)$ is a standard Gaussian random variable,
and $Q(\cdot)$ is the Q-function (the tail probability of the standard
normal distribution).\end{lem}
\begin{IEEEproof}
See Appendix \ref{sec:Miscellaneous-Proofs}.
\end{IEEEproof}

\begin{rem}
If we replace Lemma \ref{lem: output entropy for majority} in the
proof of (\ref{eq: noisy majority upper bound}) with Lemma \ref{lem: exact entropy of majority},
we can get a sharper bound than (\ref{eq: noisy majority upper bound}),
yet less explicit.
\end{rem}
In the next lemma, we evaluate $H(\maj(X^{n})|Y^{n})$ for $\alpha\approx\frac{1}{2}$. 
\begin{lem}
\label{lem:approximated entropy of majority}We have
\[
H(\maj(X^{n})|Y^{n})\geq1-\frac{1}{\pi\cdot\ln2}\left(\frac{(1-2\alpha)^{2}}{4\alpha(1-\alpha)}\right)-O\left((1-2\alpha)^{4}\right)+o(1).
\]
\end{lem}
\begin{IEEEproof}
See Appendix \ref{sec:Miscellaneous-Proofs}.
\end{IEEEproof}
We can now prove the lower bound on the SMSE of majority functions
(\ref{eq: noisy majority lower bound}).
\begin{IEEEproof}[Proof of (\ref{eq: noisy majority lower bound})]
Using Lemma \ref{lem:approximated entropy of majority} and a derivation
similar to (\ref{eq: entropy of majority}), for some $c>0$, and
all $\alpha$ sufficiently close to $\frac{1}{2}$
\begin{align}
H(Y^{n}|\maj(X^{n})) & =n-1+H(\maj(X^{n})|Y^{n})\nonumber \\
 & \geq n-\frac{1}{\pi\cdot\ln2}\left(\frac{(1-2\alpha)^{2}}{4\alpha(1-\alpha)}\right)-c\left(1-2\alpha\right)^{4}+o(1).
\end{align}
Hence, as in the proof of (\ref{eq: noiseless lower bound}) in Section
\ref{sec:The-Noiseless-Case} 
\begin{align}
\m(Y^{n}|\maj(X^{n})) & \geq\frac{n}{4}-\frac{\ln2}{2}\left[n-H(Y^{n}|\maj(X^{n}))\right]\nonumber \\
 & \geq\frac{n}{4}-\frac{1}{2\pi\alpha(1-\alpha)}\left(\frac{(1-2\alpha)^{2}}{4}\right)-c\left(1-2\alpha\right)^{4}+o(1)
\end{align}
for all sufficiently large $n$.\end{IEEEproof}
\begin{rem}
For the sake of proving (\ref{eq: noisy majority lower bound}), we
only needed the second-order approximation, given by Lemma \ref{lem:approximated entropy of majority}.
However, we note that the expression on the left-hand side of (\ref{eq: Gaussian approximation})
can be evaluated numerically to an arbitrary precision, e.g., via
a power series expansion of the analytic function $\binent\left[Q(t)\right]$.
\end{rem}

\section{Discussion and Open Problems}

The question addressed by Conjecture 1 can be equivalently cast as
an optimal sequential prediction problem, seeking the Boolean function
$\b(X^{n})$ that minimizes the cost in sequentially predicting the
channel output sequence $Y^{n}$, under logarithmic loss. Adopting
this point of view, it is natural to consider the same sequential
prediction problem under other proper loss functions. In this paper,
we have focused on the quadratic loss function. We began by considering
the noiseless case $Y^{n}=X^{n}$, which is trivial under logarithmic
loss but quite subtle under quadratic loss, and showed that majority
asymptotically achieves the minimal prediction cost among all Boolean
functions. For the case of noisy observations, we derived bounds on
the cost achievable by general Boolean functions, as well as specifically
by majority. Using these bounds, we showed that majority is better
than dictator for weak noise, but that dictator catches up and outperforms
majority for strong noise. This should be contrasted with Conjecture
\ref{conj: Boolean function conjecture}, which surmises that dictator
minimizes the sequential prediction cost under logarithmic loss, simultaneously
at all noise levels. Thus, viewed through the lens of sequential prediction,
the validity of Conjecture \ref{conj: Boolean function conjecture}
appears to possibly hinge on the unique property of logarithmic loss,
namely the fact that in the noiseless case all (balanced) Boolean
functions result in the exact same prediction cost. 

The discussion above leads us to conjecture that under quadratic loss,
there is no single sequence of functions $\{\b_{n}(X^{n})\}$ that
asymptotically minimizes the prediction cost simultaneously at all
noise levels. Moreover, it seems plausible that the optimal function
must be close to majority for weak noise, and close to dictator for
high noise. While it appears that characterizing the optimal function
at a given noise level may be difficult, it would be interesting to
understand its structural properties, e.g., whether it is monotone,
balanced, odd, etc. For logarithmic loss, it is known that the optimal
function is monotone \cite{Boolean_conjecture}. This fact can be
easily established by first switching any non-monotone coordinate
with the last coordinate (losing nothing due to the entropy chain
rule), and then \textquotedbl{}shifting\textquotedbl{} \cite{alon1983density}
the last coordinate (which can only decrease the cost, as there are
no subsequent coordinates). However, monotonicity seems more difficult
to establish under quadratic loss, even in the noiseless case; for
example, the switching/shifting technique above fails due to the lack
of a chain rule under quadratic loss. Finally, it would be interesting
to extend this study to non-Boolean functions as well as to other
proper loss functions. For example, our results readily indicate that
majority is asymptotically optimal in the noiseless case for any loss
function that behaves similarly to quadratic loss around $\frac{1}{2}$
(e.g., logarithmic loss). What is the family of proper loss functions
for which majority is asymptotically optimal?

\section*{Acknowledgment}

We are grateful to Or Ordentlich for asking the question in the noiseless
case that led to this research. We would also like to thank Or Ordentlich
and Omri Weinstein for helpful discussions, and Uri Hadar for pointing
out reference \cite{gneiting2007strictly}.

\appendices{}

\section{Miscellaneous Proofs\label{sec:Miscellaneous-Proofs}}

\subsection{Noiseless case}
\begin{IEEEproof}[Proof of Lemma \ref{lem:marginal of t-majority}]
First assume that $V^{n}$ is a $t$-majority random vector (and
not a pseudo $t$-majority random vector). From symmetry of $t$-majority
random vector, $\P\left(V_{k}=1\right)=\P\left(V_{1}=1\right)$ for
all $k\in[n]$, and so it remains to prove the statement for $k=1$.
Let us begin with the case $t\leq\frac{1}{2}$. For $t=0$ we clearly
have $P_{k}=\frac{1}{2}$. For $t=\frac{1}{2}$, the number $M_{1}$
of $\frac{1}{2}$-majority vectors such that $v_{1}=1$ ($M_{0}$
for $v_{1}=0$, respectively) is
\[
M_{1}=\sum_{m=\frac{n}{2}-1}^{n-1}{n-1 \choose m},
\]
and
\[
M_{0}=\sum_{m=\frac{n}{2}}^{n-1}{n-1 \choose m},
\]
where the index $m$ in the summation above counts the number of allowed
ones in the vector $v_{2}^{n}$. So, as $M_{1}>M_{0}$,
\[
\P\left(V_{k}=1\right)=\frac{M_{1}}{M_{0}+M_{1}}\geq\frac{1}{2}.
\]
Moreover, for all $n$ sufficiently large, 
\begin{align}
\frac{{n-1 \choose \frac{n}{2}-1}}{\sum_{m=\frac{n}{2}-1}^{n-1}{n-1 \choose m}} & \leq\frac{{n-1 \choose \frac{n}{2}-1}}{2^{n-1}}\nonumber \\
 & \trre[\leq,a]\sqrt{\frac{2}{\pi}}\cdot\frac{1}{\sqrt{n}}\cdot2^{(n-1)\left[\binent(\frac{1}{2}-\frac{1}{2(n-1)})-1\right]}\nonumber \\
 & \leq\sqrt{\frac{2}{\pi}}\cdot\frac{1}{\sqrt{n}},
\end{align}
where $(a)$ is using Lemma \ref{lem: binomial and entropy}. So
\begin{align}
\P\left(V_{k}=1\right) & =\frac{M_{1}}{M_{0}+M_{1}}\nonumber \\
 & =\frac{\sum_{m=\frac{n}{2}-1}^{n-1}{n-1 \choose m}}{\sum_{m=\frac{n}{2}}^{n-1}{n-1 \choose m}+\sum_{m=\frac{n}{2}-1}^{n-1}{n-1 \choose m}}\nonumber \\
 & =\frac{\sum_{m=\frac{n}{2}-1}^{n-1}{n-1 \choose m}}{2\cdot\sum_{m=\frac{n}{2}-1}^{n-1}{n-1 \choose m}-{n-1 \choose \frac{n}{2}-1}}\nonumber \\
 & =\frac{1}{2-\frac{{n-1 \choose \frac{n}{2}-1}}{\sum_{m=\frac{n}{2}-1}^{n-1}{n-1 \choose m}}}\nonumber \\
 & \leq\frac{1}{2}+\sqrt{\frac{2}{\pi}}\cdot\frac{1}{\sqrt{n}},
\end{align}
where in the last inequality we have used $\frac{1}{2-s}\leq\frac{1}{2}+s$,
valid for small $s$. Now, since $P_{k}$ is monotonic in $t$, then
clearly 
\[
P_{k}\leq\frac{1}{2}+\sqrt{\frac{2}{\pi}}\cdot\frac{1}{\sqrt{n}},
\]
for all $0\leq t\leq\frac{1}{2}$. 

Now for the case $t\geq\frac{1}{2}$. Using symmetry, the probability
that $V_{k}=1$ is equal to the total number of ones in the support
of $V^{n}$, divided by the total number of zeros and ones in the
support of $V^{n}$ . So, 
\[
\P\left(V_{k}=1\right)=\frac{\sum_{m=tn}^{n}{n \choose m}\cdot m}{\sum_{m=tn}^{n}{n \choose m}\cdot n}\geq\frac{\sum_{m=tn}^{n}{n \choose m}\cdot tn}{\sum_{m=tn}^{n}{n \choose m}\cdot n}\geq t.
\]
On the other hand, denoting $l_{n}\dfn n^{\nicefrac{1}{2}+\eta}$,
for all $n$ sufficiently large,
\begin{align}
\P\left(V_{k}=1\right) & =\frac{\sum_{m=tn}^{n}{n \choose m}\cdot\frac{m}{n}}{\sum_{m=tn}^{n}{n \choose m}}\nonumber \\
 & =\frac{\sum_{m=tn}^{tn+l_{n}}{n \choose m}\cdot\frac{m}{n}}{\sum_{m=tn}^{n}{n \choose m}}+\frac{\sum_{m=tn+l_{n}+1}^{n}{n \choose m}\cdot\frac{m}{n}}{\sum_{m=tn}^{n}{n \choose m}}\nonumber \\
 & \leq\frac{\sum_{m=tn}^{tn+l_{n}}{n \choose m}\cdot\left(t+\frac{l_{n}}{n}\right)}{\sum_{m=tn}^{tn+l_{n}}{n \choose m}}+\frac{\sum_{m=tn+l_{n}+1}^{n}{n \choose m}}{\sum_{m=tn}^{n}{n \choose m}}\nonumber \\
 & =t+\frac{n^{\eta}}{\sqrt{n}}+\frac{\sum_{m=tn+l_{n}+1}^{n}{n \choose m}}{\sum_{m=tn}^{n}{n \choose m}}\nonumber \\
 & \leq t+O_{\eta}\left(\frac{n^{\eta}}{\sqrt{n}}\right).
\end{align}
The last inequality follows from 
\begin{align}
\frac{\sum_{m=tn+l_{n}+1}^{n}{n \choose m}}{\sum_{m=tn}^{n}{n \choose m}} & \trre[=,a]\frac{\sum_{m=tn}^{n}{n \choose m+l_{n}+1}}{\sum_{m=tn}^{n}{n \choose m}}\nonumber \\
 & \trre[\leq,b]\max_{tn\leq m\leq n}\frac{{n \choose m+l_{n}+1}}{{n \choose m}}\nonumber \\
 & \trre[\leq,c]\max_{tn\leq m\leq n-l_{n}-1}\frac{\sqrt{8n\frac{m}{n}(1-\frac{m}{n})}}{\sqrt{\pi n\cdot\frac{m+l_{n}+1}{n}(1-\frac{m+l_{n}+1}{n})}}\cdot\frac{2^{n\binent\left(\frac{m+l_{n}+1}{n}\right)}}{2^{n\binent\left(\frac{m}{n}\right)}}\nonumber \\
 & =\left[1+o(1)\right]\sqrt{\frac{8}{\pi}}\max_{tn\leq m\leq n-l_{n}-1}2^{n\left[\binent\left(\frac{m}{n}+\frac{n^{\nicefrac{\eta}{2}}}{\sqrt{n}}\right)-\binent\left(\frac{m}{n}\right)\right]}\nonumber \\
 & \trre[\leq,d]\left[1+o(1)\right]\sqrt{\frac{8}{\pi}}\max_{\frac{n}{2}\leq m\leq n-l_{n}-1}2^{n\left[\binent\left(\frac{m}{n}+\frac{n^{\eta}}{\sqrt{n}}\right)-\binent\left(\frac{m}{n}\right)\right]}\nonumber \\
 & \trre[\leq,e]\sqrt{\frac{8}{\pi}}\cdot2^{n\left[\binent\left(\frac{1}{2}+\frac{n^{\eta}}{\sqrt{n}}\right)-\binent\left(\frac{1}{2}\right)\right]}\nonumber \\
 & \trre[\leq,f]\sqrt{\frac{8}{\pi}}\cdot2^{-\frac{2}{\ln2}n^{\eta}},
\end{align}
where $(a)$ is using the convention ${n \choose m}=0$ for $m>n$,
$(b)$ is using Lemma \ref{lem: max bound on fraction}, $(c)$ is
using Lemma \ref{lem: binomial and entropy}, $(d)$ is as $t\geq\frac{1}{2}$,
$(e)$ is because the maximum is obtained at the minimal value of
the feasible set, due the concavity of $\binent(\cdot)$, and $(f)$
is using the inequality $\binent\left(\frac{1}{2}+s\right)\leq1-\frac{2}{\ln2}s^{2}$. 

Finally, the marginal probability of $1$ for a pseudo $t$-majority
random vector is only larger than for ordinary $t$-majority random
vector, and smaller than the same marginal probability of a $(t+\frac{1}{n})$-majority
random vector. So, the asymptotic upper bound does not change for
pseudo $t$-majority random vectors.
\end{IEEEproof}

\begin{IEEEproof}[Proof of Lemma \ref{lem: entropy of partial vector}]
From the chain rule for entropies and as conditioning reduces entropy
\begin{align}
n-1 & =H(V_{1}^{n})\nonumber \\
 & =H(V_{m_{n}+1}^{n-m_{n}})+H(V_{1}^{m_{n}}|V_{m_{n}}^{n-m_{n}})+H(V_{n-m_{n}+1}^{n}|V_{1}^{n-m_{n}})\nonumber \\
 & \geq H(V_{m_{n}+1}^{n-m_{n}})+H(V_{1}^{m_{n}}|V_{m_{n}}^{n})+H(V_{n-m_{n}+1}^{n}|V_{1}^{n-m_{n}}).\label{eq: lower bound using chain rule}
\end{align}
Now, for any vector $v^{n-m_{n}}$ such that $\Hamw(v_{1}^{n-m_{n}})\geq\frac{n}{2}+1$,
it is assured that $v^{n}\in{\cal S}_{V^{n}}$, no matter what its
suffix $v_{n-m_{n}+1}^{n}$ is. Thus, conditioning on this event,
the suffix is distributed uniformly over $\{0,1\}^{m_{n}}$. This
implies that
\[
H(V_{n-m_{n}+1}^{n}|V_{1}^{n-m_{n}})\geq\P\left[\Hamw(V_{1}^{n-m_{n}})\geq\frac{n}{2}+1\right]\cdot m_{n}.
\]
Now, for all sufficiently large $n$
\begin{align}
\P\left[\Hamw(V_{1}^{n-m_{n}})\geq\frac{n}{2}+1\right] & =\frac{\sum_{k=\frac{n}{2}+1}^{n-m_{n}}{n-m_{n} \choose k}\cdot2^{m_{n}}}{2^{n-1}}\nonumber \\
 & =\frac{2\sum_{k=\frac{n}{2}+1}^{n-m_{n}}{n-m_{n} \choose k}}{2^{n-m_{n}}}\nonumber \\
 & =\frac{2\sum_{k=\frac{n-m_{n}}{2}}^{n-m_{n}}{n-m_{n} \choose k}-2\sum_{k=\frac{n-m_{n}}{2}}^{\frac{n}{2}}{n-m_{n} \choose k}}{2^{n-m_{n}}}\nonumber \\
 & \geq1-\frac{2\sum_{k=\frac{n-m_{n}}{2}}^{\frac{n}{2}}{n-m_{n} \choose k}}{2^{n-m_{n}}}\nonumber \\
 & \geq1-2\left(\frac{m_{n}}{2}+1\right)\frac{{n-m_{n} \choose \frac{n-m_{n}}{2}}}{2^{n-m_{n}}}\nonumber \\
 & \geq1-2m_{n}\frac{{n-m_{n} \choose \frac{n-m_{n}}{2}}}{2^{n-m_{n}}}\nonumber \\
 & \geq1-2\sqrt{\frac{4}{\pi(n-m_{n})}}m_{n},
\end{align}
where the last inequality is from Lemma \ref{lem: binomial and entropy}.
Recalling that $m_{n}=O(n^{\nicefrac{1}{4}-\rho})$ 
\begin{align}
H(V_{n-m_{n}+1}^{n}|V_{1}^{n-m_{n}}) & \geq m_{n}-\frac{4m_{n}^{2}}{\sqrt{\pi(n-m_{n})}}\nonumber \\
 & =m_{n}-o(1).\label{eq: bound on conditional entropy}
\end{align}
From symmetry, $H(V_{1}^{m_{n}}|V_{m_{n}}^{n-m_{n}})$ can be evaluated
to the exact same expression, and this leads to the required result.
\end{IEEEproof}

\begin{IEEEproof}[Proof of Lemma \ref{lem: exp small probability}]
Let 
\[
r_{k}\dfn\frac{(n-k+1)}{2}+(n-k+1)^{\nicefrac{1}{2}+\rho}.
\]
Then, for some $c,c'>0$
\begin{align}
\P\left[\Hamw(V_{1}^{k})\leq\frac{n}{2}-r_{k}\right] & =\P\left[\left\{ \Hamw(V_{1}^{k})\leq\frac{n}{2}-r_{k}\right\} \cap\left\{ \Hamw(V_{k+1}^{n})\geq r_{k}\right\} \right]\nonumber \\
 & \hphantom{=}+\P\left[\left\{ \Hamw(V_{1}^{k})\leq\frac{n}{2}-r_{k}\right\} \cap\left\{ \Hamw(V_{k+1}^{n})<r_{k}\right\} \right]\nonumber \\
 & =\P\left[\left\{ \Hamw(V_{1}^{k})\leq\frac{n}{2}-r_{k}\right\} \cap\left\{ \Hamw(V_{k+1}^{n})\geq r_{k}\right\} \right]\nonumber \\
 & \leq\P\left[\Hamw(V_{k+1}^{n})\geq r_{k}\right]\nonumber \\
 & \leq\frac{\sum_{l=r_{k}}^{n-k}{n-k \choose l}\cdot2^{k}}{2^{n-1}}\nonumber \\
 & \trre[\leq,a]\frac{n}{2^{n-k-1}}{n-k \choose r_{k}}\nonumber \\
 & \trre[\le,b]\frac{n}{2^{n-k-1}}2^{(n-k)\binent\left(\frac{r_{k}}{n-k}\right)}\nonumber \\
 & \trre[\le,c]2n\cdot2^{-c'(n-k)^{2\rho}}\nonumber \\
 & \leq2n\cdot2^{-c'\cdot m_{n}^{2\rho}}\nonumber \\
 & \leq2^{-c\cdot m_{n}^{2\rho}},
\end{align}
where $(a)$ is since $r_{k}\geq\frac{n-k}{2}$, $(b)$ is using Lemma
\ref{lem: binomial and entropy}, and $(c)$ is using Taylor expansion
of the binary entropy function at $\frac{1}{2}$.
\end{IEEEproof}

\subsection{Noisy case}
\begin{IEEEproof}[Proof of Lemma \ref{lem: single sample mmse over a channel}]
We have
\begin{align}
\m(W) & =\m(V+Z)\nonumber \\
 & =\m(\beta*\alpha)\nonumber \\
 & =\left[\beta(1-\alpha)+(1-\beta)\alpha\right]\cdot\left[\beta\alpha+(1-\beta)(1-\alpha)\right]\nonumber \\
 & =\alpha(1-\alpha)+(1-2\alpha)^{2}\cdot\beta(1-\beta)\nonumber \\
 & =\alpha(1-\alpha)+(1-2\alpha)^{2}\cdot\m(V).\label{eq: mmse one sample}
\end{align}

\end{IEEEproof}

\begin{IEEEproof}[Proof of Lemma \ref{lem: mmse over a channel}]
We will prove by induction. The relation holds (with equality) for
$n=1$ from Lemma \ref{lem: single sample mmse over a channel}. We
assume that the property hold up to $n-1$. Now,
\begin{align}
\m(W^{n}) & =\sum_{i=1}^{n-1}\m(W_{i}|W_{1}^{i-1})+\m(W_{n}|W_{1}^{n-1})\nonumber \\
 & \geq\sum_{i=1}^{n-1}\m(W_{i}|W_{1}^{i-1})+\m(W_{n}|W_{1}^{n-1},Z_{1}^{n-1})\nonumber \\
 & =\sum_{i=1}^{n-1}\m(W_{i}|W_{1}^{i-1})+\m(V_{n}+Z_{n}|V_{1}^{n-1},Z_{1}^{n-1})\nonumber \\
 & \trre[=,a]\sum_{i=1}^{n-1}\m(W_{i}|W_{1}^{i-1})+\m(V_{n}+Z_{n}|V_{1}^{n-1})\nonumber \\
 & \trre[=,b]\sum_{i=1}^{n-1}\m(W_{i}|W_{1}^{i-1})+\alpha(1-\alpha)+(1-2\alpha)^{2}\cdot\m(V_{n}|V_{1}^{n-1})\nonumber \\
 & \trre[\geq,c](n-1)\alpha(1-\alpha)+(1-2\alpha)^{2}\cdot\m(V_{1}^{n-1})+\alpha(1-\alpha)+(1-2\alpha)^{2}\cdot\m(V_{n}|V_{1}^{n-1})\nonumber \\
 & =n\alpha(1-\alpha)+(1-2\alpha)^{2}\cdot\m(V^{n}),\label{eq: mmse n samples}
\end{align}
where $(a)$ is since $(V_{n},Z_{n})-V_{1}^{n-1}-Z_{1}^{n-1}$, $(b)$
is using a conditional version of (\ref{eq: mmse one sample}) (which
holds since the pointwise relation holds), and $(c)$ is using the
induction assumption. Equality clearly holds when $V^{n}$ is a memoryless
random vector.
\end{IEEEproof}

\begin{IEEEproof}[Proof of Lemma \ref{lem: MMSE of beginning of vector}]
The proof is quite similar to the proof of (\ref{eq: noiseless majority})
in Section \ref{sec:The-Noiseless-Case}. Let $\rho\in(0,\nicefrac{1}{2})$
and $\eta\in[0,\frac{1}{2})$ be given. For any given $k\in[n-m_{n}]$
let us define the events 
\begin{align}
{\cal A}_{k} & \dfn\left\{ \Hamw(V_{1}^{k})\geq\frac{k-1}{2}-(n-k+1)^{\nicefrac{1}{2}+\nicefrac{\rho}{3}}\right\} \nonumber \\
 & =\left\{ \Hamw(V_{1}^{k})\geq\frac{n}{2}-r_{k}+1\right\} ,
\end{align}
where $r_{k}\dfn\frac{(n-k+1)}{2}+(n-k+1)^{\nicefrac{1}{2}+\nicefrac{\rho}{3}}$.
Let us analyze $\m(V_{k}|V_{1}^{k-1}=v_{1}^{k-1})$ for $1\leq k\leq m_{n}$
when $v_{1}^{k-1}\in{\cal A}_{k-1}$. Conditioning on $v_{1}^{k-1}\in{\cal A}_{k-1}$,
we have that $V_{k}^{n}$ is a $t$-majority vector of length $n-k+1\geq n-m_{n}+1$,
and its threshold is less than 
\[
t\leq\frac{r_{k}}{n-k+1}=\frac{1}{2}+\frac{1}{(n-k+1)^{\nicefrac{1}{2}-\nicefrac{\rho}{3}}}.
\]
Let $P_{k}\dfn\P[V_{k}=1|V_{1}^{k-1}]$. Assuming that $n$ is sufficiently
large, Lemma \ref{lem:marginal of t-majority} (with $\eta<\frac{\rho}{3}$)
implies that conditioned on the event $V^{k-1}\in{\cal A}_{k}$ 
\begin{align}
\frac{1}{2}\leq P_{k} & \leq\frac{1}{2}+\frac{1}{(n-m_{n}+1)^{\nicefrac{1}{2}-\nicefrac{\rho}{3}}}+O_{\eta}\left(\frac{1}{(n-m_{n}+1)^{\nicefrac{1}{2}-\eta}}\right)\nonumber \\
 & \leq\frac{1}{2}+O_{\eta}\left(\frac{1}{n^{\nicefrac{1}{2}-\nicefrac{\rho}{3}}}\right)
\end{align}
for all $k\in[n-m_{n}]$, and $n$ sufficiently large. Consequently,
\[
\m(V_{k}|V_{1}^{k-1}=v_{1}^{k-1})=P_{k}(1-P_{k})\geq\frac{1}{4}-O_{\eta}\left(\frac{1}{n^{1-\nicefrac{2\rho}{3}}}\right).
\]
As in Lemma \ref{lem: exp small probability} (when replacing $m_{n}$,
the maximal value of $k$, with a maximal value of $n-m_{n}$), there
exists $c>0$ such that 
\[
\P\left[V^{k-1}\not\in{\cal A}_{k-1}\right]\leq2^{-c(n-m_{n})^{\nicefrac{2\rho}{3}}}
\]
for all $k\in[m_{n}]$, and then 
\begin{align}
\sum_{k=1}^{m_{n}}\m(V_{k}|V_{1}^{k-1}) & \geq\sum_{k=1}^{m_{n}}\sum_{v^{k-1}\in{\cal A}_{k-1}}\P\left[V^{k-1}=v^{k-1}\right]\m(V_{k}|V_{1}^{k-1}=v^{k-1})\nonumber \\
 & \geq\sum_{k=1}^{m_{n}}\left[1-2^{-c(n-m_{n})^{\nicefrac{2\rho}{3}}}\right]\left[\frac{1}{4}-O_{\eta}\left(\frac{1}{n^{1-\nicefrac{2\rho}{3}}}\right)\right]\nonumber \\
 & \geq\frac{m_{n}}{4}-o_{\eta}(1).
\end{align}

\end{IEEEproof}

\begin{IEEEproof}[Proof of Lemma \ref{lem: MMSE of end of vector}]
Let us define the event 
\[
{\cal B}_{k}\dfn\left\{ \Hamw(V_{1}^{k})\geq\frac{n}{2}+1\right\} .
\]
As in the proof of Lemma \ref{lem: entropy of partial vector}, 
\begin{align}
\P\left[V^{k}\in{\cal B}_{k}\right] & \geq\P\left[\Hamw(V_{1}^{n-m_{n}})\geq\frac{n}{2}+1\right]\nonumber \\
 & \geq1-2\sqrt{\frac{4}{\pi(n-m_{n})}}m_{n}\nonumber \\
 & =1-O\left(n^{-\nicefrac{1}{4}-\rho}\right)
\end{align}
for all $k\in\{n-m_{n}+1,\ldots,n\}$. Conditioned on $v_{1}^{k-1}\in{\cal B}_{k}$,
all the suffixes $v_{k}^{n}$ are possible in order to obtain a majority
vector, and hence $\P[V_{k}=1|V_{1}^{k-1}=v_{1}^{k-1}]=\frac{1}{2}$.
Then,
\begin{align}
\sum_{k=n-m_{n}+1}^{n}\m(V_{k}|V_{1}^{k-1}) & \geq\sum_{k=n-m_{n}+1}^{n}\sum_{v_{1}^{k-1}\in{\cal B}_{k-1}}\P\left[V_{1}^{k-1}=v_{1}^{k-1}\right]\m(V_{k}|V_{1}^{k-1}=v_{1}^{k-1})\nonumber \\
 & =\sum_{k=n-m_{n}+1}^{n}\left[1-O\left(n^{-\nicefrac{1}{4}-\rho}\right)\right]\cdot\frac{1}{4}\nonumber \\
 & \geq\frac{m_{n}}{4}-O\left(\frac{1}{n^{2\rho}}\right)\nonumber \\
 & \geq\frac{m_{n}}{4}-o(1).
\end{align}

\end{IEEEproof}

\begin{IEEEproof}[Proof of Lemma \ref{lem: output entropy for majority}]
The entropy is bounded as
\begin{align}
H(Y^{n}|\maj(X^{n})=1) & \trre[=,a]H(Y^{n}|\maj(X^{n}))\nonumber \\
 & =H(\maj(X^{n})|Y^{n})+H(Y^{n})-H(\maj(X^{n}))\nonumber \\
 & =H(\maj(X^{n})|Y^{n})+n-1\nonumber \\
 & \trre[\leq,b]H(\maj(X^{n})|\maj(Y^{n}))+n-1\nonumber \\
 & \trre[\leq,c]\binent\left[\P\left(\maj(X^{n})=\maj(Y^{n})\right)\right]+n-1\nonumber \\
 & \trre[\leq,d]\mu(\alpha)+n-1+o(1),\label{eq: entropy of majority}
\end{align}
where $(a)$ follows from symmetry, $(b)$ from the data processing
theorem, $(c)$ is from Fano's inequality, and $(d)$ is from \cite[Theorem 2.45]{Bool_book}. 
\end{IEEEproof}

\begin{IEEEproof}[Proof of Lemma \ref{lem: exact entropy of majority}]
The proof of is based on the Gaussian approximation of the binomial
distribution using the Berry-Essen central limit theorem. For simplicity,
we assume that $n$ is odd, but the proof can be easily generalized
to any $n$. We begin by denoting 
\[
a(y^{n})\dfn\P[\maj(X^{n})=1|Y^{n}=y^{n}],
\]
we then writing 
\[
H(\maj(X^{n})|Y^{n})=\E\left\{ \binent\left[a(Y^{n})\right]\right\} .
\]
Since $Y^{n}$ is the output of a uniform Bernoulli random vector
$X^{n}$ through a BSC with crossover probability $\alpha$, then
$Y^{n}=X^{n}+Z^{n}$ where $Z^{n}\sim\mbox{Bern(\ensuremath{\alpha})}$.
Equivalently, we also have $X^{n}=Y^{n}+Z^{n}$, where $Y^{n}$ is
a uniform Bernoulli random vector, and $Z^{n}$ and $Y^{n}$ are independent.
We next use the Berry-Essen central limit theorem \cite[Chapter XVI.5, Theorem 2]{Feller}
to evaluate $a(y^{n})$. To this end, note that $\E[Z_{i}-\alpha]=0$,
$\E[(Z_{i}-\alpha)^{2}]=\alpha(1-\alpha)$, and $\E[|Z_{i}-\alpha|^{3}]=\alpha(1-\alpha)\left[\alpha^{2}+(1-\alpha)^{2}\right]<\infty$.
Then, 
\begin{align}
a(y^{n}) & =\P\left[\Hamw(y^{n}+Z^{n})>\frac{n}{2}\right]\nonumber \\
 & =\P\left[\sum_{i\in[n]:\; y_{i}=0}Z_{i}+\sum_{i\in[n]:\; y_{i}=1}(1-Z_{i})>\frac{n}{2}\right]\nonumber \\
 & =\P\left\{ \sum_{i\in[n]:\; y_{i}=0}(Z_{i}-\alpha)+\sum_{i\in[n]:\; y_{i}=1}(\alpha-Z_{i})>(1-2\alpha)\left[\frac{n}{2}-\Hamw(y^{n})\right]\right\} \nonumber \\
 & =\P\left\{ \frac{1}{\sqrt{n\alpha(1-\alpha)}}\left(\sum_{i\in[n]:\; y_{i}=0}(Z_{i}-\alpha)+\sum_{i\in[n]:\; y_{i}=1}(\alpha-Z_{i})\right)>\frac{(1-2\alpha)}{\sqrt{n\alpha(1-\alpha)}}\cdot\left[\frac{n}{2}-\Hamw(y^{n})\right]\right\} \nonumber \\
 & \dfn\P\left\{ S_{n}>\frac{(1-2\alpha)}{\sqrt{n\alpha(1-\alpha)}}\cdot\left[\frac{n}{2}-\Hamw(y^{n})\right]\right\} ,
\end{align}
where $S_{n}$ was implicitly defined. Now, the Berry-Essen central
limit theorem implies that for some $C_{\alpha}$
\[
\sup_{s\in\mathbb{R}}\left|\P\left[S_{n}>s\right]-\P\left[G>s\right]\right|\leq\frac{C_{\alpha}}{\sqrt{n}},
\]
where $G\sim{\cal N}(0,1)$. Further, \cite[Lemma 2.7]{csiszar2011information}
provides a bound on the difference in the entropy of two probability
distributions, in terms of the total variation distance between them.
In our case, this implies that for all $n$ sufficiently large, 
\[
\sup_{s\in\mathbb{R}}\left|\binent\left(\P\left[S_{n}>s\right]\right)-\binent\left(\P\left[G>s\right]\right)\right|\leq-\frac{2C_{\alpha}}{\sqrt{n}}\ln\left(\frac{C_{\alpha}}{\sqrt{n}}\right)=o(1).
\]
Then, denoting 
\[
H_{n}\dfn\frac{(1-2\alpha)}{\sqrt{n\alpha(1-\alpha)}}\cdot\left[\frac{n}{2}-\Hamw(y^{n})\right]
\]
we have 
\begin{align}
H(\maj(X^{n})|Y^{n}) & =\E\left\{ \binent\left[a(Y^{n})\right]\right\} \nonumber \\
 & =\E\left\{ \binent\left(\P\left[S_{n}>H_{n}\right]\right)\right\} \nonumber \\
 & =\E\left\{ \binent\left(\P\left[G>H_{n}\right]\right)\right\} +o(1)\nonumber \\
 & =\E\left\{ \binent\left[Q(|H_{n}|)\right]\right\} +o(1)
\end{align}
where $Q(\cdot)$ is the Gaussian Q-function, and in the last equality
we have used the facts that $Q(t)=1-Q(|t|)$ for $t<0$, and $\binent(p)=\binent(1-p)$.
Now, applying the central limit theorem once again, we have that $H_{n}\Rightarrow\frac{(1-2\alpha)}{\sqrt{4\alpha(1-\alpha)}}\cdot G$,
as $n\to\infty$, in distribution. To complete the proof, we note
that since $\binent\left[Q(|t|)\right]$ is a bounded and continuous
function of $t$, Portmanteau's lemma (e.g. \cite[Chapter VIII.1, Theorem 1]{Feller})
implies that 
\[
\E\left\{ \binent\left[Q(|H_{n}|)\right]\right\} \to\E\left\{ \binent\left[Q\left(\frac{|(1-2\alpha)G|}{\sqrt{4\alpha(1-\alpha)}}\right)\right]\right\} ,
\]
as $n\to\infty$, concluding the proof.
\end{IEEEproof}

\begin{IEEEproof}[Proof of Lemma \ref{lem:approximated entropy of majority}]
Let us denote $\alpha=\frac{1}{2}-\gamma$ for $\gamma\in(0,\frac{1}{2})$,
and then let us inspect 
\[
\E\left\{ \binent\left[Q(\Gamma)\right]\right\} \dfn\E\left\{ \binent\left[Q\left(\frac{\left|G\right|\gamma}{\sqrt{(\frac{1}{2}-\gamma)(\frac{1}{2}+\gamma)}}\right)\right]\right\} 
\]
as $\gamma\downarrow0$. Using Leibniz's integral rule, we obtain
$Q'(t)=-\frac{1}{\sqrt{2\pi}}e^{-\nicefrac{t^{2}}{2}}$, $Q''(t)=\frac{t}{\sqrt{2\pi}}\cdot e^{-\nicefrac{t^{2}}{2}}$
and so, there exists $\overline{c}>0$ such that for all $t\geq0$
\[
Q(t)\geq\frac{1}{2}-\frac{t}{\sqrt{2\pi}}.
\]
Similarly, there exists $\tilde{c},s_{1}>0$ such that for all $s\in(0,s_{1})$
 
\[
\binent\left(\frac{1}{2}-s\right)\geq1-\frac{2}{\ln2}s^{2}-\tilde{c}s^{4}.
\]
Hence, for all sufficiently small $t>0$
\begin{align}
\binent[Q(t)] & =\binent\left[\frac{1}{2}-\left(\frac{1}{2}-Q(t)\right)\right]\nonumber \\
 & \geq1-\frac{2}{\ln2}\left(\frac{1}{2}-Q(t)\right)^{2}-\tilde{c}\left(\frac{1}{2}-Q(t)\right)^{4}\nonumber \\
 & \geq1-\frac{1}{\pi\cdot\ln2}t^{2}-\frac{\tilde{c}}{4\pi^{2}}t^{4}.
\end{align}
So, there exists $\hat{c}>0$ such that for all sufficiently small
$\gamma$, 
\begin{align}
 & \hphantom{=}\E\left\{ \binent\left[Q\left(\Gamma\right)\right]\right\} \nonumber \\
 & \geq\P\left[\left|G\right|\leq\gamma^{-1+\rho}\right]\cdot\E\left\{ \binent\left[Q\left(\Gamma\right)\right]\vert\left|G\right|\leq\gamma^{-1+\rho}\right\} \nonumber \\
 & \geq\P\left[\left|G\right|\leq\gamma^{-1+\rho}\right]\cdot\E\left\{ 1-\frac{1}{\pi\cdot\ln2}\Gamma^{2}-\frac{\tilde{c}}{4\pi^{2}}\Gamma^{4}\vert\left|G\right|\leq\gamma^{-1+\rho}\right\} \nonumber \\
 & =\int_{-\gamma^{-1+\rho}}^{\gamma^{-1+\rho}}\frac{1}{\sqrt{2\pi}}e^{-\nicefrac{t^{2}}{2}}\cdot\left[1-\frac{1}{\pi\cdot\ln2}\left(\frac{\gamma^{2}t^{2}}{(\frac{1}{2}-\gamma)(\frac{1}{2}+\gamma)}\right)-\frac{\tilde{c}}{4\pi^{2}}\left(\frac{\gamma^{4}t^{4}}{(\frac{1}{2}-\gamma)^{2}(\frac{1}{2}+\gamma)^{2}}\right)\right]\cdot dt\nonumber \\
 & =1-2Q(\gamma^{-1+\rho})-\int_{-\gamma^{-1+\rho}}^{\gamma^{-1+\rho}}\frac{1}{\sqrt{2\pi}}e^{-\nicefrac{t^{2}}{2}}\cdot\left[\frac{1}{\pi\cdot\ln2}\left(\frac{\gamma^{2}t^{2}}{(\frac{1}{2}-\gamma)(\frac{1}{2}+\gamma)}\right)+\frac{\tilde{c}}{4\pi^{2}}\left(\frac{\gamma^{4}t^{4}}{(\frac{1}{2}-\gamma)^{2}(\frac{1}{2}+\gamma)^{2}}\right)\right]\cdot dt\nonumber \\
 & \geq1-2Q(\gamma^{-1+\rho})-\int_{-\infty}^{\infty}\frac{1}{\sqrt{2\pi}}e^{-\nicefrac{t^{2}}{2}}\cdot\left[\frac{1}{\pi\cdot\ln2}\left(\frac{\gamma^{2}t^{2}}{(\frac{1}{2}-\gamma)(\frac{1}{2}+\gamma)}\right)+\frac{\tilde{c}}{4\pi^{2}}\left(\frac{\gamma^{4}t^{4}}{(\frac{1}{2}-\gamma)^{2}(\frac{1}{2}+\gamma)^{2}}\right)\right]\cdot dt\nonumber \\
 & =1-2Q(\gamma^{-1+\rho})-\frac{1}{\pi\cdot\ln2}\left(\frac{\gamma^{2}}{(\frac{1}{2}-\gamma)(\frac{1}{2}+\gamma)}\right)-\frac{\tilde{c}}{4\pi^{2}}\left(\frac{3\gamma^{4}}{(\frac{1}{2}-\gamma)^{2}(\frac{1}{2}+\gamma)^{2}}\right)\nonumber \\
 & \trre[\geq,a]1-\frac{1}{\pi\cdot\ln2}\left(\frac{\gamma^{2}}{(\frac{1}{2}-\gamma)(\frac{1}{2}+\gamma)}\right)-\hat{c}\gamma^{4},
\end{align}
where $(a)$ is since for any $\rho\in(0,1)$, using $Q(t)\leq\frac{1}{t}\cdot e^{-\nicefrac{t^{2}}{2}}$
we have 
\begin{equation}
\P\left[\left|G\right|\geq\gamma^{-1+\rho}\right]=2Q(\gamma^{-1+\rho})\leq2\gamma^{1-\rho}\cdot\exp\left(-\frac{1}{2\gamma^{2-2\rho}}\right).\label{eq: upper bound on Q function}
\end{equation}

\end{IEEEproof}

\section{Useful Results}
\begin{lem}[{\cite[Lemma 17.5.1]{Cover:2006:EIT:1146355}}]
\label{lem: binomial and entropy} For $0<\alpha<1$ such that $n\alpha$
is integer
\[
\frac{2^{n\binent(\alpha)}}{\sqrt{8n\alpha(1-\alpha)}}\leq{n \choose n\alpha}\leq\frac{2^{n\binent(\alpha)}}{\sqrt{\pi n\alpha(1-\alpha)}}.
\]

\end{lem}

\begin{lem}[{\cite[Lemma 1]{cover1996universal}}]
\label{lem: max bound on fraction}If $\{a_{i}\}_{i=1}^{n}$ and
$\{b_{i}\}_{i=1}^{n}$ are all non-negative numbers, then
\[
\frac{\sum_{i=1}^{n}a_{i}}{\sum_{i=1}^{n}b_{i}}\leq\max_{1\leq i\leq n}\frac{a_{i}}{b_{i}}.
\]
\end{lem}
\begin{cor}
Under the conditions above and for any integer $l>0$,
\[
\frac{\sum_{i=1}^{n-l}a_{i}}{\sum_{i=1}^{n}b_{i}}\leq\max_{1\leq i\leq n-l}\frac{a_{i}}{b_{i}}.
\]
This can be obtained by replacing $a_{i}$ with $0$ for $n-l+1\leq i\leq n$.
\end{cor}
\bibliographystyle{plain}
\bibliography{Predictable_Half_Space}

\end{document}